\documentclass[superscriptaddress,
twocolumn,
bibnotes,amsmath,amssymb,aps,pra,floatfix]{revtex4-2}

\usepackage[T1]{fontenc}
\usepackage[latin9]{inputenc}
\usepackage{graphicx}
\usepackage{hyperref}
\usepackage{esint}
\usepackage{times}
\usepackage{helvet}
\usepackage{braket}
\usepackage{bbm}
\usepackage{xcolor}
\usepackage{enumitem}
\usepackage[toc,page,titletoc]{appendix}

\newcommand{\JILA}{JILA, NIST, and Department of Physics, University of Colorado, Boulder, CO 80309, USA}
\newcommand{\CTQM}{Center for Theory of Quantum Matter, University of Colorado, Boulder, CO 80309, USA}
\newcommand{\Toronto}{Department of Physics and CQIQC, University of Toronto, Ontario M5S~1A7, Canada}

\begin{document}

\title{Collective P-Wave Orbital Dynamics of Ultracold Fermions}
\date{\today}

\author{Mikhail Mamaev}
\email{mikhail.mamaev@colorado.edu}
\affiliation{\JILA}
\affiliation{\CTQM}
\author{Peiru He}
\affiliation{\JILA}
\affiliation{\CTQM}
\author{Thomas Bilitewski}
\affiliation{\JILA}
\affiliation{\CTQM}
\author{Vijin Venu}
\affiliation{\Toronto}
\author{Joseph H. Thywissen}
\affiliation{\Toronto}
\author{Ana Maria Rey}
\affiliation{\JILA}
\affiliation{\CTQM}

\begin{abstract}
{
We consider the non-equilibrium orbital dynamics of spin-polarized ultracold fermions in the first excited band of an optical lattice. A specific lattice depth and filling configuration is designed to allow the $p_x$ and $p_y$ excited orbital degrees of freedom to act as a pseudo-spin.
Starting from the full Hamiltonian for p-wave interactions in a periodic potential, we derive an extended Hubbard-type model that describes the anisotropic lattice dynamics of the excited orbitals at low energy. We then show how dispersion engineering can provide a viable route to realizing collective behavior driven by p-wave interactions. In particular, Bragg dressing and lattice depth can reduce single-particle dispersion rates, such that a collective many-body gap is opened with only moderate Feshbach enhancement of p-wave interactions. Physical insight into the emergent gap-protected collective dynamics is gained by projecting the Hamiltonian into the Dicke manifold, yielding a one-axis twisting model for the orbital pseudo-spin that can be probed using conventional Ramsey-style interferometry. Experimentally realistic protocols to prepare and measure the many-body dynamics are discussed, including the effects of band relaxation, particle loss, spin-orbit coupling, and doping.
}
\end{abstract}
\maketitle

{\it Introduction.}
Ultracold quantum gases in optical lattices are among the leading platforms for quantum simulation of strongly correlated matter and  non-equilibrium dynamics. While there has been impressive experimental progress~\cite{gross2017quantum,schafer2020review}, most investigations thus far have been limited to s-wave interacting systems in the lowest motional band. A fascinating avenue yet to be explored experimentally is many-body lattice physics with p-wave interactions~\cite{gurarie2005pWave,gurarie2007pWave} in higher bands. P-wave interacting systems can host long-sought phases including topological superfluids, Majorana fermions~\cite{elliott2020topological,read2000fqh,levinsen2007pwave}, and itinerant ferromagnetism~\cite{hui2015ferro,jiang2016ferro,yang2016magnetism,kurlov2019magnetism,singh2020ferromagnetism}. At the same time, atoms in higher bands are a unique resource~\cite{noh2016quantum} for emulating orbital degrees of freedom in real materials~\cite{tokura2000orbital} which give rise to heavy fermions~\cite{coleman2007heavyFermions}, RKKY interactions~\cite{ruderman1954rkky}, and orbitally ordered Mott phases~\cite{imada1998insulator,khaliullin2005insulator}.

Despite these attractive features, control and manipulation of p-wave interacting gases has remained a challenge for ultracold atom experiments. The timescales on which p-wave interactions contribute to dynamics tend to be slow compared to coherence times~\cite{martin2013olc} and lossy when increased by a Feshbach resonance~\cite{demarco1999feshbach,regal2003feshbach,luciuk2016feshbach}. Moreover, collisions in higher bands suffer from band relaxation~\cite{spielman2006collision,muller2007orbital}. Important progress in mitigating relaxation has been made via designed lattice geometries~\cite{wirth2011orbitalSuperfluidity,kock2015orbitalSuperfluidity,diliberto2016orbitalSuperfluidity} and symmetry protection~\cite{hartke2021dfsFermions}, but further advances are required to explore the full range of orbital physics in optical lattices.

Here we consider the problem of non-equilibrium orbital physics in an optical lattice, and identify a limit in which collective dynamics emerge. Orbital dynamics in first excited bands are stabilized via Pauli blocking by preparing a spin-polarized system with a completely filled ground band, mimicking the conventional conduction-band configuration of materials. P-wave interactions are enabled by the orbitally antisymmetric two-atom wavefunctions. We explore the use of Bragg dressing to suppress orbital anisotropy, which allows an accurate mapping of the p-wave Fermi-Hubbard model to an XXZ spin model, in which the conventional magnetic spin states are replaced by orbital states. We delineate a specific regime in which the collective dynamics can be further mapped to a collective one-axis twisting (OAT) model thanks to the opening of a many-body gap~\cite{rey2008gap}. Dispersion engineering lowers the demands on Feshbach-tuned interaction strength, and thus elastic interactions can dominate over inelastic collisions and other decoherence processes. We further discuss how the p-wave induced mean-field dynamics can be observed with a Ramsey protocol. 

The conceptual map that we develop offers new ways to understand p-wave orbital physics in an experimentally accessible regime. We connect previously established real-space pseudo-potential formulations to a tractable extended Fermi-Hubbard model, and use laser driving as a tool to coordinate interaction-driven dynamics. We show that a simple collective model can explain the emergent gap-protected dynamics.

{\it P-wave Fermi-Hubbard model.}
The scenario we consider is a three-dimensional (3D) optical lattice loaded with spin-polarized fermionic  atoms in their ground electronic state. The system Hamiltonian can be written in terms of field operators $\hat{\psi}(\vec{R})$ acting in  real space $\vec{R}=(X,Y,Z)$ as
\begin{equation}
\begin{aligned}
    \hat{H} &= \int d^3\vec{R}\>\> \hat{\psi}^{\dagger} \bigg[-\frac{\hbar^2}{2m}\vec{\nabla}^2+ \sum_{\nu=X,Y,Z} V_{\nu}E_r\sin^2\left(\frac{\pi \nu}{a}\right)\bigg]\hat{\psi}\\
    &- \sum_{\nu={X,Y,Z}}\frac{3\pi \hbar^2b_{\nu}^3}{2m}\int d^{3}\vec{R} \>\>  
    W_\nu (\hat{\psi}^{\dagger},\hat{\psi}^{\dagger})W_\nu (\hat{\psi},\hat{\psi}),
\end{aligned}
\end{equation}
where $W_\nu(\hat{A},\hat{B})=(\vec{\nabla}_{\nu}\hat{A}) \hat{B} - \hat{A}  (\vec{\nabla}_{\nu}\hat{B})$. The first line includes the kinetic energy and lattice potential, where the lattice depth along $\nu \in \{X,Y,Z\}$ is $V_{\nu} E_r$ with $E_r$ the recoil energy, $a$ is the lattice spacing and $m$ is the atomic mass. We assume that $V_{X}= V_{Y} \ll V_{Z}$, confining the system to independent 2D planes. The second line contains the collisional interactions, which are p-wave since the s-wave channel is blocked for a spin-polarized gas. We have used a pseudo-potential approximation~\cite{Pricoupenko:2006cj,Idziaszek:2006gz,Idziaszek:2009jq,Zinner:2012gi} with two different scattering volumes due to dipole-dipole splitting of the closed channel, $b_{X}^3=b_{Y}^3\equiv b_{XY}^3$ and $b_{Z}^3$, controlled by a p-wave Feshbach resonance. We assume a magnetic field pointed along the tight confined direction $Z$; for such a field, we will show that only the transverse volume $b_{XY}^3$ is relevant for the interactions that our specific configuration will exhibit.
We also note that while the p-wave scattering volume tends to have strong energy dependence, we operate in the regime where the energy dependence can be well approximated as an additional constant shift in the position of the Feshbach resonance~\cite{SM}.\nocite{Idziaszek:2009jq,Chiu:2018fl,swallows2011spinwaveMatrixElements}

\begin{figure}[t!]
\centering
\includegraphics[width=1.0\linewidth]{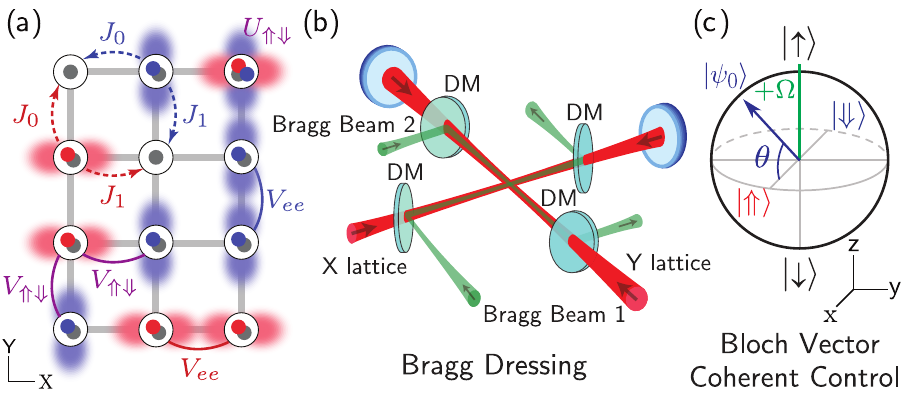}
\caption{Conceptual schematic. (a) Fermi-Hubbard physics on a single $X$-$Y$ plane. The $\Uparrow$ ($X$-excited) and $\Downarrow$ ($Y$-excited) atoms tunnel at rates $J_0$ and $J_1$ along their ground and excited directions respectively. There is an on-site p-wave interaction $U_{\Uparrow\Downarrow}$ between $\Uparrow$, $\Downarrow$ atoms, as well as nearest-neighbour interactions $V_{ee}$, $V_{\Uparrow\Downarrow}$.
(b) Bragg dressing coupling $\Uparrow$, $\Downarrow$ can be implemented with beams (shown in green) that co-propagate with the lattice beams (red), when the Bragg-laser wavelength is half that of the lattice beams. The out-of-plane lattice beam is not shown.
(c) Effective Bloch sphere of the Bragg-dressed spin states. The $\Uparrow$, $\Downarrow$ states are equal superpositions of the two flavors of the dressed basis. Using standard coherent control protocols, any direction of the Bloch vector can be initialized.
}
\label{fig_Schematic}
\end{figure}

We seek to work in the ultracold regime where atoms only occupy the ground band and the first-excited bands of the $X$ and $Y$ directions, with orbitals $\phi^\alpha_{\vec{r}} (\vec{R})$ given by 
\begin{equation}
\begin{aligned}
\phi^g_{\vec{r}} (\vec{R}) &= w_0^{X}(X-ia ) w_0^{Y}(Y-ja) w_0^{Z}(Z),\\
\phi^\Uparrow_{\vec{r}} (\vec{R}) &= w_1^{X}(X-ia ) w_0^{Y }(Y-ja) w_0^{Z}(Z), \,\mbox{and}\\
\phi^\Downarrow_{\vec{r}} (\vec{R}) &= w_0^{X}(X-ia) w_1^{Y}(Y-ja) w_0^{Z}(Z),\\
\end{aligned}
\end{equation}
where $\vec{r}=\{ia,ja\}$ is the lattice position on the 2D plane and $w_n^{\nu}(\nu- i a )$ is the $n$-th lattice Wannier function localized at site $i$ of direction $\nu$. Our desired configuration is a filling of $N/L$=2: each site will have one atom in the $g$ orbital, filling the ground band, and a second atom in the degenerate subspace of the excited orbitals $\{\Uparrow, \Downarrow\}$ acting as a spin-1/2 degree of freedom. The occupation of the ground state prevents collisional relaxation since for any energy-conserving two-atom process, leaving the excited subspace would require an atom to move down to the ground band; here, this is forbidden by Pauli exclusion.

The low-energy Hamiltonian can be written as an anisotropic multi-orbital  model  by projecting $\hat{H}$ into the Wannier basis defined by the three chosen orbital states, yielding
\begin{equation}
\label{eq_FermiModel}
    \hat{H}_{\mathrm{FH}} = \hat{H}_{J} + \hat{H}_{\mathrm{int}}.
\end{equation}
Here $\hat{H}_J$ describes the kinetic energy of the excited atoms, which tunnel to nearest neighbour sites at rate $J_0$ or $J_1$ depending on the tunneling direction and orbital: the $\Uparrow$ atoms tunnel at rate $J_1$ along $X$ and rate $J_0$ along $Y$, while the $\Downarrow$ atoms do the opposite [see Fig.~\ref{fig_Schematic}(a)]. In general $J_1 \gg J_0$ since excited states have a larger spatial extent along their excitation direction. Since the $g$ atoms are in a filled band, they are Pauli blocked and do not contribute to $\hat{H}_{J}$. For the excited atoms, the tunneling Hamiltonian can be written in momentum space as
\begin{equation}
\begin{aligned}
\hat{H}_{J} &= \sum_{\vec{k}} \epsilon_{\vec{k}} \left(\hat{n}_{\vec{k},\Uparrow} - \hat{n}_{\vec{k},\Downarrow}\right)+\sum_{\vec{k}} \bar{E}_{\vec{k}} \left(\hat{n}_{\vec{k},\Uparrow} + \hat{n}_{\vec{k},\Downarrow}\right),
\end{aligned}
\end{equation}
with $\epsilon_{\vec{k}} = (J_1 + J_0)[\cos (k_Xa) - \cos (k_Ya)]$ and $\bar{E}_{\vec{k}} = (J_1 - J_0)[\cos (k_Xa) + \cos (k_Ya)]$. Here $\hat{n}_{\vec{k},\gamma} = \hat{c}_{\vec{k},\gamma}^{\dagger}\hat{c}_{\vec{k},\gamma}$ and $\hat{c}_{\vec{k},\gamma} = L^{-1/2}\sum_{\vec{r}}e^{i \vec{r} \cdot \vec{k}}\hat{c}_{\vec{r},\gamma}$, with $\vec{k} = (k_X, k_Y)$ the lattice quasi-momentum and $\hat{c}_{\vec{r},\gamma}$ annihilating an atom on lattice site $\vec{r}$ in band state $\gamma \in \{\Uparrow,\Downarrow\}$.

The second term $\hat{H}_{\mathrm{int}}$ contains the interactions. Using the Wannier expansion these take the form of
\begin{eqnarray}
&&    \hat{H}_{\mathrm{int}} \approx \sum_{\vec{r}, \vec{r}', \vec{r}'', \vec{r}'''}\sum_{\alpha,\beta,\sigma,\gamma} U_{\vec{r}, \vec{r}', \vec{r}'', \vec{r}'''}^{\alpha\beta\sigma\gamma} \hat{c}_{\vec{r},\alpha}^{\dagger}\hat{c}_{\vec{r}',\beta}^{\dagger}\hat{c}_{\vec{r}'',\sigma}\hat{c}_{\vec{r}''',\gamma},\\
&&U_{\vec{r}, \vec{r}', \vec{r}'', \vec{r}'''}^{\alpha\beta\sigma\gamma}=G^{XY} \sum_{\nu=X,Y}\int d^{3}\vec{R} W_\nu(\phi_{\vec{r}}^{\alpha*},\phi_{\vec{r}'}^{\beta*})W_\nu(\phi_{\vec{r}''}^\sigma,\phi_{\vec{r}'''}^\gamma) ,\notag
\end{eqnarray}
where $G^{XY}=-\frac{3\pi\hbar^2 b_{XY}^3}{2m}$,  $\{\vec{r},\vec{r}',\vec{r}'',\vec{r}'''\}$ each run over all lattice sites and  $\{\alpha,\beta,\sigma ,\gamma\}$ over band states $\{g,\Uparrow,\Downarrow\}$. Since all included orbitals have the same wavefunction along the $Z$ direction and we assume tight confinement $V_{Z}$ restricting the system to 2D planes, only contributions from the terms proportional to the transverse volume $ b_{XY}^3$ are relevant.

We evaluate all these terms, and keep the ones that have non-negligible coefficient $U_{\vec{r}, \vec{r}', \vec{r}'', \vec{r}'''}^{\alpha\beta\sigma\gamma}$ on realistic timescales and are not inhibited by a band gap or another stronger interaction~\cite{SM}. For a sufficiently deep lattice $V_{X}=V_{Y} \gg 1$ the relevant terms give rise to an extended Fermi-Hubbard model which consists of on-site (OS) and nearest-neighbour (NN) interactions, $\hat{H}_{\mathrm{int}}= \hat{H}_{\mathrm{int}}^{\mathrm{(OS)}} + \hat{H}_{\mathrm{int}}^{\mathrm{(NN)}}$. The on-site term is
\begin{equation}
\label{eq_PWaveInteractionsOnSite}
\hat{H}_{\mathrm{int}}^{\mathrm{(OS)}}\approx U_{\Uparrow\Downarrow}\sum_{\vec{r}} \hat{n}_{\vec{r},\Uparrow}\hat{n}_{\vec{r},\Downarrow},
\end{equation}
corresponding to a density-density interaction between $\Uparrow$ and $\Downarrow$ atoms with strength $U_{\Uparrow\Downarrow}=
4U_{\vec{r}, \vec{r}, \vec{r}, \vec{r}}^{\Uparrow\Downarrow\Downarrow\Uparrow}$. On-site interactions between $\Uparrow$, $g$ and between $\Downarrow$, $g$ are also present, but amount to a constant of motion for $V_{X}=V_{Y}$ and can be dropped. The nearest-neighbour terms are anisotropic density-density interactions given by 
\begin{equation}
\label{eq_PWaveInteractionsCrossSite}
\begin{aligned}
\hat{H}_{\mathrm{int}}^{\mathrm{(NN)}}\approx &V_{ee}\sum_{\vec{r}}\left(  \hat{n}_{\vec{r},\Uparrow}\hat{n}_{\vec{r}+\vec{r}_{X},\Uparrow}+\hat{n}_{\vec{r},\Downarrow}\hat{n}_{\vec{r}+\vec{r}_{Y},\Downarrow}\right)\\
+&V_{\Uparrow\Downarrow}\sum_{\vec{r},\nu=X,Y}\left(\hat{n}_{\vec{r},\Uparrow}\hat{n}_{\vec{r}+\vec{r}_{\nu},\Downarrow}+\hat{n}_{\vec{r},\Downarrow}\hat{n}_{\vec{r}+\vec{r}_{\nu},\Uparrow}\right).
\end{aligned}
\end{equation}
Here $\vec{r}_{\nu}$ is a lattice unit vector along the $\nu \in \{X, Y\}$ direction. The interaction $V_{ee}=4U_{\vec{r}, \vec{r}+ \vec{r}_{X}, \vec{r}+ \vec{r}_{X},  \vec{r}}^{\Uparrow\Uparrow\Uparrow\Uparrow}$ is between nearest-neighbour pairs of atoms both in the same excited orbital along their excitation direction, as depicted in Fig.~\ref{fig_Schematic}(a). $V_{\Uparrow\Downarrow}=4U_{\vec{r}, \vec{r}+ \vec{r}_{X\Downarrow}, \vec{r}+ \vec{r}_{X\Downarrow},  \vec{r}}^{\Uparrow\Downarrow\Downarrow\Uparrow}$ is an interaction between nearest neighbour atoms in different excited orbitals. For a sample atom choice of $^{40}$K and parameters of $V_X$=$V_Y$=25, $V_Z$=100, $b_{XY}$=292$a_0$ with $a_0$ the Bohr radius (a 20-fold increase in background volume), we predict coefficients of $J_0$=5Hz, $J_1$=130Hz, $U_{\Uparrow\Downarrow}$=900Hz, $V_{ee}$=0.3Hz, $V_{\Uparrow\Downarrow}$=0.1Hz. These parameters are used in the following calculations, unless otherwise specified.

{\it Momentum-space spin model.}
The implementation of an anisotropic extended Fermi-Hubbard model, Eq.~\eqref{eq_FermiModel} already offers exciting opportunities for quantum simulation~\cite{baier2016extended}. However, as a first step we are specifically interested in regimes amenable for theoretical analysis, starting from a fully polarized initial  state, where nevertheless p-wave interactions play a dominant role. For our p-wave system, however, the large spin dependent dispersion in  $\hat{H}_{J}$ will induce fast single particle dynamics that quickly depolarizes the initial state. To favor ordering of the orbital states, one can reduce competitive depolarization via the introduction of a laser field that couples $\Uparrow$ and $\Downarrow$:
\begin{equation}
\label{eq_Drive}
    \hat{H}_{\Omega} = \frac{\Omega}{2}\sum_{\vec{k}}\left(\hat{c}_{\vec{k},\Uparrow}^{\dagger}\hat{c}_{\vec{k},\Downarrow}+h.c.\right).
\end{equation}
Experimentally, such a term can be generated by an optical field whose Bragg grating is oriented along a diagonal reciprocal lattice vector [see Fig.~\ref{fig_Schematic}(b)]. We assume that the drive couples only atoms with equal quasi-momentum, which can be ensured with appropriate laser wavelengths and orientation~\cite{SM}. Dressed with this coupling, the single-particle eigenenergies $E^{\pm}_{\vec{k}}$ of the atoms change from $\overline{E}_{\vec{k}}\pm \epsilon_{\vec{k}}$ to $\overline{E}_{\vec{k}}\pm\sqrt{\epsilon_{\vec{k}}^2+(\Omega/2)^2 }$. When $\Omega/2 \gg |\epsilon_{\vec{k}}|$, the anisotropic part of the spectrum $\epsilon_{\vec{k}}$ is flattened, which allows interactions to play a more dominant role in the spin dynamics.

Under the assumption of a strong drive $\Omega/2 \gg |\epsilon_{\vec{k}}|$, the flattened spectrum suppresses quasi-momentum-changing collisions between the atoms, which renders each atom frozen in a given $\vec{k}$-mode when evolving from a collective initial product state. In this regime, also known as the collisionless regime~\cite{Laloe,Rey2014}, we can approximate the Fermi-Hubbard model with a spin-1/2 model $\hat{H}_{\mathrm{FH}}+\hat{H}_{\Omega}\approx \hat{H}_{S}$:
\begin{equation}
\label{eq_SpinModel}
\hat{H}_{\mathrm{S}}=\sum_{\vec{k},\vec{k}'}U_{\vec{k},\vec{k}'}\vec{\sigma}_{\vec{k}}\cdot\vec{\sigma}_{\vec{k}'}+\sum_{\vec{k},\vec{k}'}V_{\vec{k},\vec{k}'}\hat{\sigma}_{\vec{k}}^{x}\hat{\sigma}_{\vec{k}'}^{x}+\sum_{\vec{k}}\left(\epsilon_{\vec{k}}\hat{\sigma}_{\vec{k}}^{x}+\frac{\Omega}{2}\hat{\sigma}_{\vec{k}}^{z}\right),\notag
\end{equation}
with coefficients
\small
\begin{equation}
\begin{aligned}
    U_{\vec{k},\vec{k}'}&=-\frac{U_{\Uparrow\Downarrow}}{4L}-\frac{V_{\Uparrow\Downarrow}}{2L}\left[\cos(k_X a- k_X'a)+\cos(k_Ya-k_Y'a)\right],\\
    V_{\vec{k},\vec{k}'}&=\frac{V_{ee}-2V_{\Uparrow\Downarrow}}{4L}\left[2-\cos(k_X a - k_X'a)-\cos(k_Ya-k_Y'a)\right].\notag
\end{aligned}
\end{equation}
\normalsize
Here we define spin operators $\hat{\sigma}^{\alpha}_{\vec{k}} =\hat{a}_{\vec{k},\mu}^{\dagger}\sigma^{\alpha}_{\mu\mu'}\hat{a}_{\vec{k},\mu'}$, with $\sigma^{\nu}$ the standard $2\times2$ Pauli matrices for $\alpha \in \{x,y,z\}$, summing over new dressed atom flavors $\mu,\mu' \in \{\uparrow,\downarrow\}$ that are eigenstates of the drive [see Fig.~\ref{fig_Schematic}(c)]:
\begin{equation}
\label{eq_DressedFermiOperators}
\hat{a}_{\vec{k},\uparrow} = \frac{1}{\sqrt{2}}(\hat{c}_{\vec{k},\Uparrow} + \hat{c}_{\vec{k},\Downarrow}) \>\>\>\mbox{and} \>\>\>\hat{a}_{\vec{k},\downarrow} = \frac{1}{\sqrt{2}}(\hat{c}_{\vec{k},\Uparrow} - \hat{c}_{\vec{k},\Downarrow}).
\end{equation}
The on-site contribution proportional to $U_{\Uparrow\Downarrow}$ is SU(2) symmetric, because only the orbital singlet state of the two excited bands can interact, while the nearest-neighbour terms yield XXZ-type anisotropicity.

{\it Ramsey spectroscopy.} To probe the system dynamics, we consider time-evolution of a collective product state
\begin{equation}
\label{eq_TiltedInitialState}
    \ket{\psi_0} = e^{i \theta \hat{S}^{y}} \prod_{\vec{k}} \ket{\rightarrow}_{\vec{k}},
\end{equation}
where $\ket{\rightarrow}_{\vec{k}}=(\ket{\uparrow}_{\vec{k}}+\ket{\downarrow}_{\vec{k}})/ \sqrt{2}$ is an $X$-excited ($\Uparrow$) band state and $\hat{S}^{\alpha = x,y,z} = \frac{1}{2}\sum_{\vec{k}}\hat{\sigma}_{\vec{k}}^{\alpha=x,y,z}$ are collective-spin operators. This state corresponds to either all spins pointing along the $x$ direction of the dressed Bloch sphere, or inclined at some angle $\theta$ into the $x-z$ plane [see Fig.~\ref{fig_Schematic}(c)]. Such a state can be prepared from a band insulator by using Raman coupling schemes and control over the lattice depth~\cite{SM}. We still assume ideal filling of 2 atoms per site, although a small hole fraction can be tolerated~\cite{SM}.

To probe the dynamics of this initial state we propose a Ramsey-style protocol. The system is initialized and evolved for a time $t/2$ under the full Hamiltonian. The sign of the drive is then quenched from $+\Omega \to - \Omega$ with e.g. a fast pulse of the laser detuning, and the system is evolved for another time $t/2$, undoing the drive's single-particle rotation. Then the collective observable $\langle\hat{S}^{+}\rangle=\langle\hat{S}_{x}\rangle+i\langle\hat{S}_{y}\rangle \equiv C(t) e^{i \phi(t)}$ is measured where $C = (\langle \hat{S}^{x}\rangle^2 + \langle \hat{S}^{y}\rangle^2)^{1/2}$ is the contrast, and $\phi = \text{arg} \langle \hat{S}^{+} \rangle$ an interaction-induced phase shift.

Measurements of such collective spin observables are straightforward to implement as the excited bands have different spatial distributions upon being released from the lattice. Turning off both the drive and the lattice and measuring the resulting gas cloud's $X$-band population ($\Uparrow$) via band-mapping~\cite{kohl2005bandMapping} allows measurements of $\langle \hat{S}^{x}\rangle$. Leaving the drive on for an additional time $t\Omega = \pi/2$ after the Ramsey protocol rotates $y$ into $x$, allowing the measurement of $\langle \hat{S}^{y} \rangle$ via an $\langle \hat{S}^{x} \rangle$ measurement. While $\langle \hat{S}^{z} \rangle$ is in principle conserved for $\Omega/2 \gg |\epsilon_{\vec{k}}|$, we can also measure it by advancing the relative phase of the Bragg beams ahead by $\pi/2$, which allows us to use the drive for a $\pi/2$ pulse that rotates $z$ into $x$, and then measuring $\langle \hat{S}^{x} \rangle$ once more.

\begin{figure}[t!]
\centering
\includegraphics[width=1.0\linewidth]{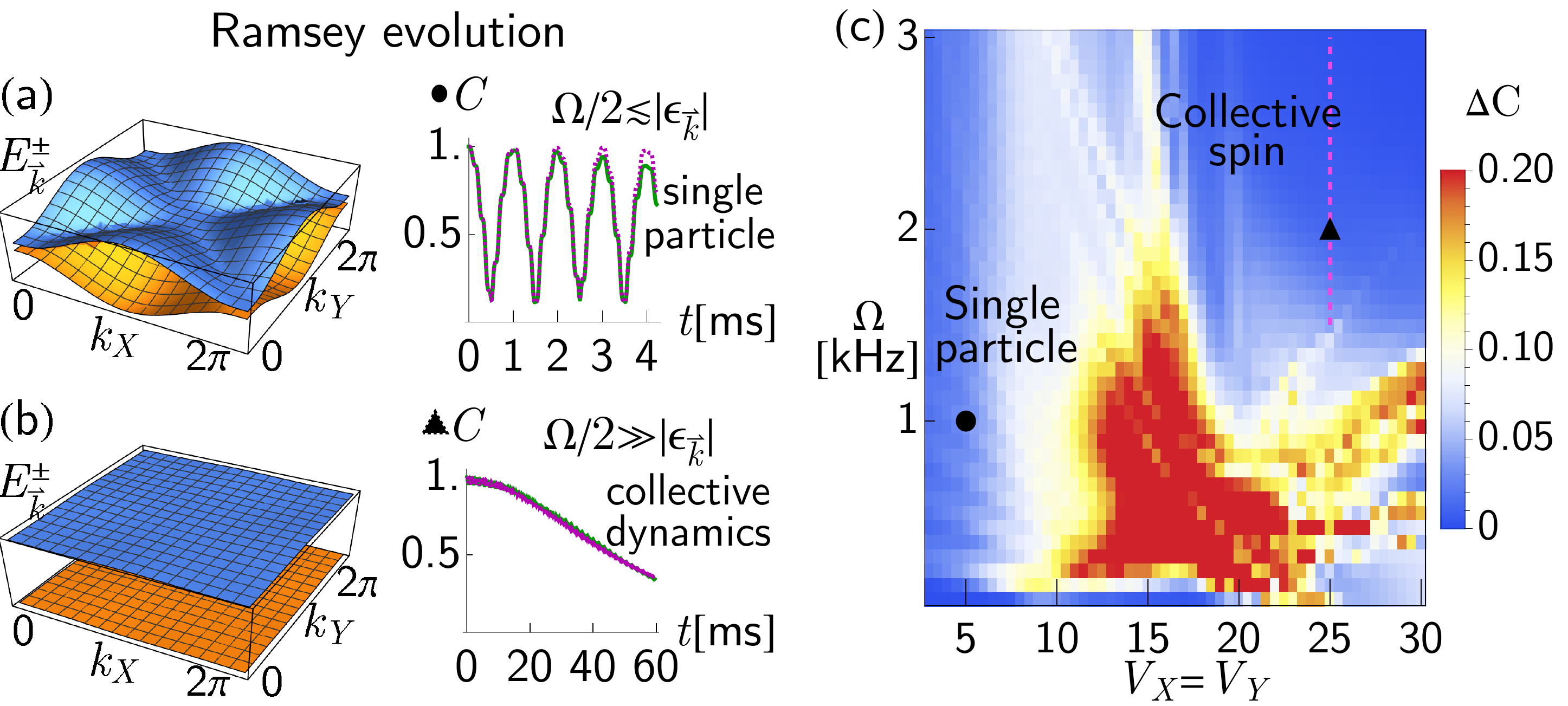}
\caption{(a-b) Single-particle spectrum $E^{\pm}_{\vec{k}} = \overline{E}_{\vec{k}}\pm \sqrt{\epsilon_{\vec{k}}^2 + (\Omega/2)^2}$ and characteristic contrast time-evolution for (a) a weak drive $\Omega/2 \lesssim |\epsilon_{\vec{k}}|$ and (b) a strong drive $\Omega/2 \gg |\epsilon_{\vec{k}}|$, for the Fermi-Hubbard+drive model $\hat{H}_{\mathrm{FH}}+\hat{H}_{\Omega}$ (green) and spin model $\hat{H}_{\mathrm{S}}$ (purple). (c) Benchmark comparison of the two models' agreement. Both models are evolved from a product state $\ket{\psi_0}$ with $\theta = 0$ to a fixed time $t_f = 50/J_1$, and their contrast $C$ is compared with a root-mean-square error $\Delta C = (\frac{1}{t_f}\int_0^{t_f} dt |\frac{2}{L}\left(C_{\mathrm{S}}-C_{\mathrm{FH+\Omega}}\right)|^2 )^{1/2}$, truncated to $\text{min}(\Delta C, 0.2)$ for clarity, using a small system $L=3\times 2$. The representative evolutions in panels (a-b) are indicated by the circle and triangle respectively. The purple dashed line indicates the collective regime explored further in Fig.~\ref{fig_Evolution}.}
\label{fig_SpinValidity}
\end{figure}

Figures~\ref{fig_SpinValidity}(a-b) show the single-particle spectrum $E^{\pm}_{\vec{k}}$ and representative time-evolution of the contrast for both the driven Fermi-Hubbard model $\hat{H}_{\mathrm{FH}}+\hat{H}_{\Omega}$ and the spin model $\hat{H}_{\mathrm{S}}$, starting from $\ket{\psi_0}$ and setting $\theta = 0$. Panel (a) corresponds to the case of a weak drive $\Omega/2 \lesssim |\epsilon_{\vec{k}}|$ and (b) to the case of a strong drive $\Omega/2 \gg |\epsilon_{\vec{k}}|$. We see a characteristic crossover from fast single-particle dynamics to a slow collective interaction-induced decay. To more clearly identify these regimes and benchmark our spin model mapping, we compare the time evolution of the two models in Fig.~\ref{fig_SpinValidity}(c) with a root mean square error of the contrast. The spin model is valid when either the lattice depth is very shallow and single-particle tunneling dominates, or when the drive is strong enough to flatten the spectrum and make the single particle dispersion subdominant with respect to the p-wave exchange interactions. At this point a many-body gap energetically suppresses single-particle dynamics and keeps the spins aligned, allowing for collective behaviour~\cite{smale2019gapProtectionDPT,rey2008gap}.

{\it One-axis twisting.} When in the collective,  gap-protected regime, the dominant spin model terms are the Heisenberg term $-\frac{U_{\Uparrow\Downarrow}}{4L} \sum_{\vec{k},\vec{k}'}\vec{\sigma}_{\vec{k}}\cdot \vec{\sigma}_{\vec{k}'}$ and the drive $\frac{\Omega}{2} \sum_{\vec{k}}\hat{\sigma}_{\vec{k}}^{z}$. Both these terms conserve the total spin $S$, defined by $\vec{S}\cdot\vec{S}\ket{S,M} = S(S+1)\ket{S,M}$ where  $\vec{S}=(\hat{S}^{x},\hat{S}^{y},\hat{S}^z)$, and $\ket{S,M}$ are collective-spin eigenstates with non-negative $S\in \frac{L}{2}, \frac{L}{2}-1,\dots$ and projection $M \in S, S-1,\dots,-S$ (satisfying $\hat{S}^{z}\ket{S,M}=M \ket{S,M}$). A spin-polarize  initial state in the fully-symmetric Dicke manifold $S = L/2$ will be confined to that manifold, as transitions to other manifolds induced by the kinetic terms will be energetically suppressed by the many-body gap~\cite{smale2019gapProtectionDPT,chu2020gap,rey2008gap}. This permits us to further simplify the Hamiltonian by projecting it into the Dicke manifold~\cite{SM}, yielding $\hat{H}_{S} \approx \hat{H}_{\mathrm{OAT}}$, where
\begin{equation}
\label{eq_CollectiveModel}
    \hat{H}_{\mathrm{OAT}} = -({U_{\Uparrow\Downarrow}}/{L}) \vec{S}\cdot \vec{S}+\chi \hat{S}^{z}\hat{S}^{z}+\Omega \hat{S}^{z}.
\end{equation}
This is a one-axis twisting (OAT) model, which is well studied for its entanglement generation in the form of spin squeezing~\cite{kitagawa1993squeezed}. The coefficient $\chi$ is
\begin{equation}
\begin{aligned}
    \chi = \frac{1}{L-1}\frac{2(J_0+J_1)^2U_{\Uparrow\Downarrow}}{\Omega^2 - U_{\Uparrow\Downarrow}^2}- \frac{1}{L}\left(V_{ee}-2V_{\Uparrow\Downarrow}\right).
\end{aligned}
\end{equation}
The first term comes from the tunneling, and the second from the nearest-neighbour interactions.

\begin{figure}[tb!]
\centering
\includegraphics[width=1.0\linewidth]{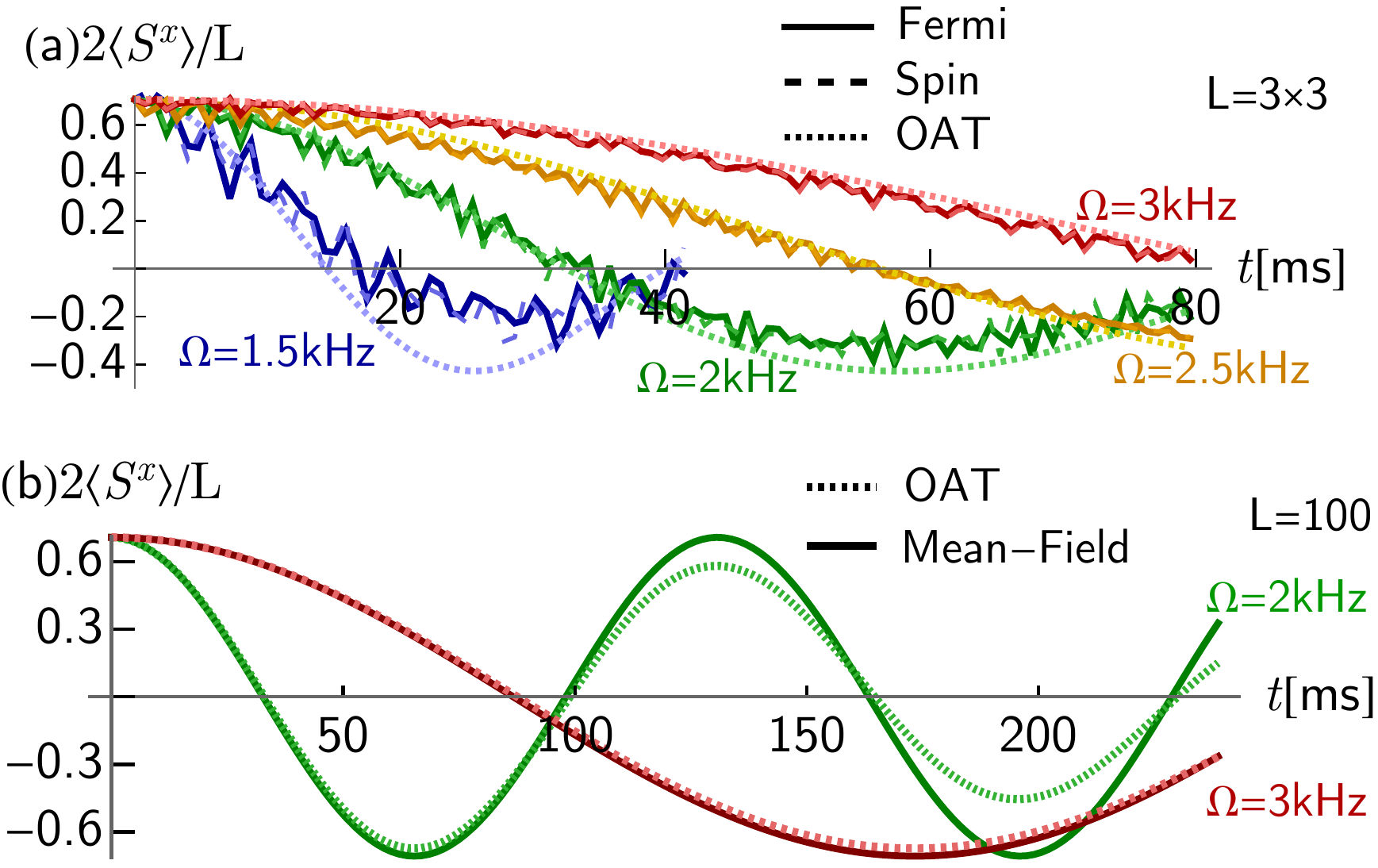}
\caption{(a) Time evolution of $\langle\hat{S}^{x}\rangle = C \cos(\phi)$ to measure the density phase shift $\phi$, comparing the Fermi-Hubbard model+drive, $\hat{H}_{\mathrm{FH}}+\hat{H}_{\Omega}$, spin model $\hat{H}_{\mathrm{S}}$ and one-axis twisting model $\hat{H}_{\mathrm{OAT}}$, for system size $L=3\times 3$ and inclination angle $\theta = \pi/4$. The parameters used lie along the purple dashed line in the previous Fig.~\ref{fig_SpinValidity}(c). (b) Time-evolution of $\langle \hat{S}^{x}\rangle$ for a larger system of $L=100$, using only the OAT model together with its predicted mean-field behaviour.}
\label{fig_Evolution}
\end{figure}

The coefficient $\chi$ can be measured using our Ramsey protocol through the phase shift $\phi$. At the mean-field level, under the OAT model the collective spin rotates about the $z$ axis of the Bloch sphere at a rate $\langle\hat{S}^{+}\rangle = \frac{L}{2}e^{i\phi(t)},$ with $\phi(t)= 2 \chi \langle \hat{S}^{z}\rangle t=\chi L \sin(\theta) t$ where $\langle \hat{S}^{z}\rangle = \frac{L}{2}\sin(\theta)$ is conserved. Fig.~\ref{fig_Evolution}(a) shows sample time-evolutions of $\langle \hat{S}^{x}\rangle=\text{Re}[\langle \hat{S}^{+}\rangle]$ with a tilt angle $\theta = \pi/4$ for both the OAT and the underlying Fermi-Hubbard and spin models. We see the expected oscillation with period set by $2 \chi \langle \hat{S}^{z} \rangle$. In Fig.~\ref{fig_Evolution}(b) we show the same dynamics for a larger system using only the OAT. The frequency of the oscillations is not very sensitive to system size since $\chi\sim 1/L$ and $\langle\hat{S}^{z}\rangle \sim L \sin(\theta)$. Since the amplitude of the oscillations is proportional to the contrast, which decays more slowly with increasing $L$, better visibility of the oscillations is possible in larger systems. 

{\it Conclusions and Outlook.}
We have shown a robust and experimentally realistic protocol for observing long-sought p-wave physics in optical lattices. Our specific band configuration and laser dressing allows one to isolate the interaction dynamics via collective enhancement, and see a signal on realistic timescales without the usual challenges of band relaxation or losses due to strong Feshbach resonance. The system can be reduced to a simple one-axis twisting model described by a single interaction parameter $\chi$, which is straightforward to measure while capturing the dominant many-body p-wave effects.

While in this work we focus on simple dynamics probed via mean-field Ramsey spectroscopy, well controlled spin interactions such as OAT provide avenues to useful many-body entanglement generation and non-equilibrium quantum simulation. Further progress can realize more complex and interesting extended Fermi-Hubbard models~\cite{dutta2015extendedHubbard} that are theoretically challenging, and yet straightforward to implement in experiment using extensions of our basic scheme. This system also allows the exploration of non-collective physics, including pairing, the effects of vacancies, or local quantum correlations, using tools such as quantum gas microscopes already implemented in several state-of-the-art optical lattice experiments~\cite{parsons2015microscope,haller2015microscope,cheuk2015microscope,omran2015microscope,edge2015microscope}.

\textbf{Acknowledgements.} We thank Leo Radzihovsky and John Bohn for their useful feedback and careful reading of our manuscript. This work is supported by the AFOSR grants FA9550-19-1-0275, FA9550-19-1-7044, and FA9550-19-1-0365, by ARO W911NF-15-1-0603, by the NSF JILA-PFC PHY-1734006 grant, by NIST, and by NSERC.

\bibliography{PWaveBiblio.bib}

\begin{thebibliography}{49}%
\makeatletter
\providecommand \@ifxundefined [1]{%
 \@ifx{#1\undefined}
}%
\providecommand \@ifnum [1]{%
 \ifnum #1\expandafter \@firstoftwo
 \else \expandafter \@secondoftwo
 \fi
}%
\providecommand \@ifx [1]{%
 \ifx #1\expandafter \@firstoftwo
 \else \expandafter \@secondoftwo
 \fi
}%
\providecommand \natexlab [1]{#1}%
\providecommand \enquote  [1]{``#1''}%
\providecommand \bibnamefont  [1]{#1}%
\providecommand \bibfnamefont [1]{#1}%
\providecommand \citenamefont [1]{#1}%
\providecommand \href@noop [0]{\@secondoftwo}%
\providecommand \href [0]{\begingroup \@sanitize@url \@href}%
\providecommand \@href[1]{\@@startlink{#1}\@@href}%
\providecommand \@@href[1]{\endgroup#1\@@endlink}%
\providecommand \@sanitize@url [0]{\catcode `\\12\catcode `\$12\catcode
  `\&12\catcode `\#12\catcode `\^12\catcode `\_12\catcode `\%12\relax}%
\providecommand \@@startlink[1]{}%
\providecommand \@@endlink[0]{}%
\providecommand \url  [0]{\begingroup\@sanitize@url \@url }%
\providecommand \@url [1]{\endgroup\@href {#1}{\urlprefix }}%
\providecommand \urlprefix  [0]{URL }%
\providecommand \Eprint [0]{\href }%
\providecommand \doibase [0]{https://doi.org/}%
\providecommand \selectlanguage [0]{\@gobble}%
\providecommand \bibinfo  [0]{\@secondoftwo}%
\providecommand \bibfield  [0]{\@secondoftwo}%
\providecommand \translation [1]{[#1]}%
\providecommand \BibitemOpen [0]{}%
\providecommand \bibitemStop [0]{}%
\providecommand \bibitemNoStop [0]{.\EOS\space}%
\providecommand \EOS [0]{\spacefactor3000\relax}%
\providecommand \BibitemShut  [1]{\csname bibitem#1\endcsname}%
\let\auto@bib@innerbib\@empty
\bibitem [{\citenamefont {Gross}\ and\ \citenamefont
  {Bloch}(2017)}]{gross2017quantum}%
  \BibitemOpen
  \bibfield  {author} {\bibinfo {author} {\bibfnamefont {C.}~\bibnamefont
  {Gross}}\ and\ \bibinfo {author} {\bibfnamefont {I.}~\bibnamefont {Bloch}},\
  }\bibfield  {title} {\bibinfo {title} {Quantum simulations with ultracold
  atoms in optical lattices},\ }\href@noop {} {\bibfield  {journal} {\bibinfo
  {journal} {Science}\ }\textbf {\bibinfo {volume} {357}},\ \bibinfo {pages}
  {995} (\bibinfo {year} {2017})}\BibitemShut {NoStop}%
\bibitem [{\citenamefont {Sch{\"a}fer}\ \emph {et~al.}(2020)\citenamefont
  {Sch{\"a}fer}, \citenamefont {Fukuhara}, \citenamefont {Sugawa},
  \citenamefont {Takasu},\ and\ \citenamefont {Takahashi}}]{schafer2020review}%
  \BibitemOpen
  \bibfield  {author} {\bibinfo {author} {\bibfnamefont {F.}~\bibnamefont
  {Sch{\"a}fer}}, \bibinfo {author} {\bibfnamefont {T.}~\bibnamefont
  {Fukuhara}}, \bibinfo {author} {\bibfnamefont {S.}~\bibnamefont {Sugawa}},
  \bibinfo {author} {\bibfnamefont {Y.}~\bibnamefont {Takasu}},\ and\ \bibinfo
  {author} {\bibfnamefont {Y.}~\bibnamefont {Takahashi}},\ }\bibfield  {title}
  {\bibinfo {title} {Tools for quantum simulation with ultracold atoms in
  optical lattices},\ }\href@noop {} {\bibfield  {journal} {\bibinfo  {journal}
  {Nat. Rev. Phys.}\ }\textbf {\bibinfo {volume} {2}},\ \bibinfo {pages} {411}
  (\bibinfo {year} {2020})}\BibitemShut {NoStop}%
\bibitem [{\citenamefont {Gurarie}\ \emph {et~al.}(2005)\citenamefont
  {Gurarie}, \citenamefont {Radzihovsky},\ and\ \citenamefont
  {Andreev}}]{gurarie2005pWave}%
  \BibitemOpen
  \bibfield  {author} {\bibinfo {author} {\bibfnamefont {V.}~\bibnamefont
  {Gurarie}}, \bibinfo {author} {\bibfnamefont {L.}~\bibnamefont
  {Radzihovsky}},\ and\ \bibinfo {author} {\bibfnamefont {A.}~\bibnamefont
  {Andreev}},\ }\bibfield  {title} {\bibinfo {title} {{Quantum phase
  transitions across a p-wave Feshbach resonance}},\ }\href@noop {} {\bibfield
  {journal} {\bibinfo  {journal} {Phys. Rev. Lett.}\ }\textbf {\bibinfo
  {volume} {94}},\ \bibinfo {pages} {230403} (\bibinfo {year}
  {2005})}\BibitemShut {NoStop}%
\bibitem [{\citenamefont {Gurarie}\ and\ \citenamefont
  {Radzihovsky}(2007)}]{gurarie2007pWave}%
  \BibitemOpen
  \bibfield  {author} {\bibinfo {author} {\bibfnamefont {V.}~\bibnamefont
  {Gurarie}}\ and\ \bibinfo {author} {\bibfnamefont {L.}~\bibnamefont
  {Radzihovsky}},\ }\bibfield  {title} {\bibinfo {title} {Resonantly paired
  fermionic superfluids},\ }\href@noop {} {\bibfield  {journal} {\bibinfo
  {journal} {Ann. Phys.}\ }\textbf {\bibinfo {volume} {322}},\ \bibinfo {pages}
  {2} (\bibinfo {year} {2007})}\BibitemShut {NoStop}%
\bibitem [{\citenamefont {Elliott}\ and\ \citenamefont
  {Franz}(2015)}]{elliott2020topological}%
  \BibitemOpen
  \bibfield  {author} {\bibinfo {author} {\bibfnamefont {S.~R.}\ \bibnamefont
  {Elliott}}\ and\ \bibinfo {author} {\bibfnamefont {M.}~\bibnamefont
  {Franz}},\ }\bibfield  {title} {\bibinfo {title} {{Colloquium: Majorana
  fermions in nuclear, particle, and solid-state physics}},\ }\href
  {https://doi.org/10.1103/RevModPhys.87.137} {\bibfield  {journal} {\bibinfo
  {journal} {Rev. Mod. Phys.}\ }\textbf {\bibinfo {volume} {87}},\ \bibinfo
  {pages} {137} (\bibinfo {year} {2015})}\BibitemShut {NoStop}%
\bibitem [{\citenamefont {Read}\ and\ \citenamefont
  {Green}(2000)}]{read2000fqh}%
  \BibitemOpen
  \bibfield  {author} {\bibinfo {author} {\bibfnamefont {N.}~\bibnamefont
  {Read}}\ and\ \bibinfo {author} {\bibfnamefont {D.}~\bibnamefont {Green}},\
  }\bibfield  {title} {\bibinfo {title} {{Paired states of fermions in two
  dimensions with breaking of parity and time-reversal symmetries and the
  fractional quantum Hall effect}},\ }\href
  {https://doi.org/10.1103/PhysRevB.61.10267} {\bibfield  {journal} {\bibinfo
  {journal} {Phys. Rev. B}\ }\textbf {\bibinfo {volume} {61}},\ \bibinfo
  {pages} {10267} (\bibinfo {year} {2000})}\BibitemShut {NoStop}%
\bibitem [{\citenamefont {Levinsen}\ \emph {et~al.}(2007)\citenamefont
  {Levinsen}, \citenamefont {Cooper},\ and\ \citenamefont
  {Gurarie}}]{levinsen2007pwave}%
  \BibitemOpen
  \bibfield  {author} {\bibinfo {author} {\bibfnamefont {J.}~\bibnamefont
  {Levinsen}}, \bibinfo {author} {\bibfnamefont {N.~R.}\ \bibnamefont
  {Cooper}},\ and\ \bibinfo {author} {\bibfnamefont {V.}~\bibnamefont
  {Gurarie}},\ }\bibfield  {title} {\bibinfo {title} {Strongly resonant
  $p$-wave superfluids},\ }\href
  {https://doi.org/10.1103/PhysRevLett.99.210402} {\bibfield  {journal}
  {\bibinfo  {journal} {Phys. Rev. Lett.}\ }\textbf {\bibinfo {volume} {99}},\
  \bibinfo {pages} {210402} (\bibinfo {year} {2007})}\BibitemShut {NoStop}%
\bibitem [{\citenamefont {Hui}\ \emph {et~al.}(2015)\citenamefont {Hui},
  \citenamefont {Brydon}, \citenamefont {Sau}, \citenamefont {Tewari},\ and\
  \citenamefont {Sarma}}]{hui2015ferro}%
  \BibitemOpen
  \bibfield  {author} {\bibinfo {author} {\bibfnamefont {H.-Y.}\ \bibnamefont
  {Hui}}, \bibinfo {author} {\bibfnamefont {P.}~\bibnamefont {Brydon}},
  \bibinfo {author} {\bibfnamefont {J.~D.}\ \bibnamefont {Sau}}, \bibinfo
  {author} {\bibfnamefont {S.}~\bibnamefont {Tewari}},\ and\ \bibinfo {author}
  {\bibfnamefont {S.~D.}\ \bibnamefont {Sarma}},\ }\bibfield  {title} {\bibinfo
  {title} {Majorana fermions in ferromagnetic chains on the surface of bulk
  spin-orbit coupled s-wave superconductors},\ }\href@noop {} {\bibfield
  {journal} {\bibinfo  {journal} {Sci. Rep.}\ }\textbf {\bibinfo {volume}
  {5}},\ \bibinfo {pages} {8880} (\bibinfo {year} {2015})}\BibitemShut
  {NoStop}%
\bibitem [{\citenamefont {Jiang}\ \emph {et~al.}(2016)\citenamefont {Jiang},
  \citenamefont {Kurlov}, \citenamefont {Guan}, \citenamefont {Schreck},\ and\
  \citenamefont {Shlyapnikov}}]{jiang2016ferro}%
  \BibitemOpen
  \bibfield  {author} {\bibinfo {author} {\bibfnamefont {Y.}~\bibnamefont
  {Jiang}}, \bibinfo {author} {\bibfnamefont {D.~V.}\ \bibnamefont {Kurlov}},
  \bibinfo {author} {\bibfnamefont {X.-W.}\ \bibnamefont {Guan}}, \bibinfo
  {author} {\bibfnamefont {F.}~\bibnamefont {Schreck}},\ and\ \bibinfo {author}
  {\bibfnamefont {G.~V.}\ \bibnamefont {Shlyapnikov}},\ }\bibfield  {title}
  {\bibinfo {title} {{Itinerant ferromagnetism in one-dimensional two-component
  Fermi gases}},\ }\href {https://doi.org/10.1103/PhysRevA.94.011601}
  {\bibfield  {journal} {\bibinfo  {journal} {Phys. Rev. A}\ }\textbf {\bibinfo
  {volume} {94}},\ \bibinfo {pages} {011601} (\bibinfo {year}
  {2016})}\BibitemShut {NoStop}%
\bibitem [{\citenamefont {Yang}\ \emph {et~al.}(2016)\citenamefont {Yang},
  \citenamefont {Guan},\ and\ \citenamefont {Cui}}]{yang2016magnetism}%
  \BibitemOpen
  \bibfield  {author} {\bibinfo {author} {\bibfnamefont {L.}~\bibnamefont
  {Yang}}, \bibinfo {author} {\bibfnamefont {X.}~\bibnamefont {Guan}},\ and\
  \bibinfo {author} {\bibfnamefont {X.}~\bibnamefont {Cui}},\ }\bibfield
  {title} {\bibinfo {title} {Engineering quantum magnetism in one-dimensional
  trapped fermi gases with p-wave interactions},\ }\href@noop {} {\bibfield
  {journal} {\bibinfo  {journal} {Physical Review A}\ }\textbf {\bibinfo
  {volume} {93}},\ \bibinfo {pages} {051605} (\bibinfo {year}
  {2016})}\BibitemShut {NoStop}%
\bibitem [{\citenamefont {Kurlov}\ \emph {et~al.}(2019)\citenamefont {Kurlov},
  \citenamefont {Matveenko}, \citenamefont {Gritsev},\ and\ \citenamefont
  {Shlyapnikov}}]{kurlov2019magnetism}%
  \BibitemOpen
  \bibfield  {author} {\bibinfo {author} {\bibfnamefont {D.}~\bibnamefont
  {Kurlov}}, \bibinfo {author} {\bibfnamefont {S.}~\bibnamefont {Matveenko}},
  \bibinfo {author} {\bibfnamefont {V.}~\bibnamefont {Gritsev}},\ and\ \bibinfo
  {author} {\bibfnamefont {G.}~\bibnamefont {Shlyapnikov}},\ }\bibfield
  {title} {\bibinfo {title} {One-dimensional two-component fermions with
  contact even-wave repulsion and su (2)-symmetry-breaking near-resonant
  odd-wave attraction},\ }\href@noop {} {\bibfield  {journal} {\bibinfo
  {journal} {Physical Review A}\ }\textbf {\bibinfo {volume} {99}},\ \bibinfo
  {pages} {043631} (\bibinfo {year} {2019})}\BibitemShut {NoStop}%
\bibitem [{\citenamefont {Singh}\ \emph {et~al.}(2020)\citenamefont {Singh},
  \citenamefont {Pilati},\ and\ \citenamefont
  {Orso}}]{singh2020ferromagnetism}%
  \BibitemOpen
  \bibfield  {author} {\bibinfo {author} {\bibfnamefont {M.}~\bibnamefont
  {Singh}}, \bibinfo {author} {\bibfnamefont {S.}~\bibnamefont {Pilati}},\ and\
  \bibinfo {author} {\bibfnamefont {G.}~\bibnamefont {Orso}},\ }\bibfield
  {title} {\bibinfo {title} {Itinerant ferromagnetism in the repulsive hubbard
  chain with spin-anisotropic odd-wave attraction},\ }\href@noop {} {\bibfield
  {journal} {\bibinfo  {journal} {Physical Review A}\ }\textbf {\bibinfo
  {volume} {102}},\ \bibinfo {pages} {053301} (\bibinfo {year}
  {2020})}\BibitemShut {NoStop}%
\bibitem [{\citenamefont {Noh}\ and\ \citenamefont
  {Angelakis}(2016)}]{noh2016quantum}%
  \BibitemOpen
  \bibfield  {author} {\bibinfo {author} {\bibfnamefont {C.}~\bibnamefont
  {Noh}}\ and\ \bibinfo {author} {\bibfnamefont {D.~G.}\ \bibnamefont
  {Angelakis}},\ }\bibfield  {title} {\bibinfo {title} {Quantum simulations and
  many-body physics with light},\ }\href@noop {} {\bibfield  {journal}
  {\bibinfo  {journal} {Rep. Prog. Phys.}\ }\textbf {\bibinfo {volume} {80}},\
  \bibinfo {pages} {016401} (\bibinfo {year} {2016})}\BibitemShut {NoStop}%
\bibitem [{\citenamefont {Tokura}\ and\ \citenamefont
  {Nagaosa}(2000)}]{tokura2000orbital}%
  \BibitemOpen
  \bibfield  {author} {\bibinfo {author} {\bibfnamefont {Y.}~\bibnamefont
  {Tokura}}\ and\ \bibinfo {author} {\bibfnamefont {N.}~\bibnamefont
  {Nagaosa}},\ }\bibfield  {title} {\bibinfo {title} {Orbital physics in
  transition-metal oxides},\ }\href@noop {} {\bibfield  {journal} {\bibinfo
  {journal} {Science}\ }\textbf {\bibinfo {volume} {288}},\ \bibinfo {pages}
  {462} (\bibinfo {year} {2000})}\BibitemShut {NoStop}%
\bibitem [{\citenamefont {Coleman}(2007)}]{coleman2007heavyFermions}%
  \BibitemOpen
  \bibfield  {author} {\bibinfo {author} {\bibfnamefont {P.}~\bibnamefont
  {Coleman}},\ }\bibinfo {title} {Heavy fermions: {E}lectrons at the edge of
  magnetism},\ in\ \href
  {https://doi.org/https://doi.org/10.1002/9780470022184.hmm105} {\emph
  {\bibinfo {booktitle} {{Handbook of Magnetism and Advanced Magnetic
  Materials}}}}\ (\bibinfo {year} {2007})\BibitemShut {NoStop}%
\bibitem [{\citenamefont {Ruderman}\ and\ \citenamefont
  {Kittel}(1954)}]{ruderman1954rkky}%
  \BibitemOpen
  \bibfield  {author} {\bibinfo {author} {\bibfnamefont {M.~A.}\ \bibnamefont
  {Ruderman}}\ and\ \bibinfo {author} {\bibfnamefont {C.}~\bibnamefont
  {Kittel}},\ }\bibfield  {title} {\bibinfo {title} {Indirect exchange coupling
  of nuclear magnetic moments by conduction electrons},\ }\href
  {https://doi.org/10.1103/PhysRev.96.99} {\bibfield  {journal} {\bibinfo
  {journal} {Phys. Rev.}\ }\textbf {\bibinfo {volume} {96}},\ \bibinfo {pages}
  {99} (\bibinfo {year} {1954})}\BibitemShut {NoStop}%
\bibitem [{\citenamefont {Imada}\ \emph {et~al.}(1998)\citenamefont {Imada},
  \citenamefont {Fujimori},\ and\ \citenamefont {Tokura}}]{imada1998insulator}%
  \BibitemOpen
  \bibfield  {author} {\bibinfo {author} {\bibfnamefont {M.}~\bibnamefont
  {Imada}}, \bibinfo {author} {\bibfnamefont {A.}~\bibnamefont {Fujimori}},\
  and\ \bibinfo {author} {\bibfnamefont {Y.}~\bibnamefont {Tokura}},\
  }\bibfield  {title} {\bibinfo {title} {Metal-insulator transitions},\ }\href
  {https://doi.org/10.1103/RevModPhys.70.1039} {\bibfield  {journal} {\bibinfo
  {journal} {Rev. Mod. Phys.}\ }\textbf {\bibinfo {volume} {70}},\ \bibinfo
  {pages} {1039} (\bibinfo {year} {1998})}\BibitemShut {NoStop}%
\bibitem [{\citenamefont {Khaliullin}(2005)}]{khaliullin2005insulator}%
  \BibitemOpen
  \bibfield  {author} {\bibinfo {author} {\bibfnamefont {G.}~\bibnamefont
  {Khaliullin}},\ }\bibfield  {title} {\bibinfo {title} {{Orbital order and
  fluctuations in Mott insulators}},\ }\href@noop {} {\bibfield  {journal}
  {\bibinfo  {journal} {Progress of Theoretical Physics Supplement}\ }\textbf
  {\bibinfo {volume} {160}},\ \bibinfo {pages} {155} (\bibinfo {year}
  {2005})}\BibitemShut {NoStop}%
\bibitem [{\citenamefont {Martin}\ \emph {et~al.}(2013)\citenamefont {Martin},
  \citenamefont {Bishof}, \citenamefont {Swallows}, \citenamefont {Zhang},
  \citenamefont {Benko}, \citenamefont {Von-Stecher}, \citenamefont {Gorshkov},
  \citenamefont {Rey},\ and\ \citenamefont {Ye}}]{martin2013olc}%
  \BibitemOpen
  \bibfield  {author} {\bibinfo {author} {\bibfnamefont {M.}~\bibnamefont
  {Martin}}, \bibinfo {author} {\bibfnamefont {M.}~\bibnamefont {Bishof}},
  \bibinfo {author} {\bibfnamefont {M.}~\bibnamefont {Swallows}}, \bibinfo
  {author} {\bibfnamefont {X.}~\bibnamefont {Zhang}}, \bibinfo {author}
  {\bibfnamefont {C.}~\bibnamefont {Benko}}, \bibinfo {author} {\bibfnamefont
  {J.}~\bibnamefont {Von-Stecher}}, \bibinfo {author} {\bibfnamefont
  {A.}~\bibnamefont {Gorshkov}}, \bibinfo {author} {\bibfnamefont
  {A.}~\bibnamefont {Rey}},\ and\ \bibinfo {author} {\bibfnamefont
  {J.}~\bibnamefont {Ye}},\ }\bibfield  {title} {\bibinfo {title} {A quantum
  many-body spin system in an optical lattice clock},\ }\href@noop {}
  {\bibfield  {journal} {\bibinfo  {journal} {Science}\ }\textbf {\bibinfo
  {volume} {341}},\ \bibinfo {pages} {632} (\bibinfo {year}
  {2013})}\BibitemShut {NoStop}%
\bibitem [{\citenamefont {DeMarco}\ \emph {et~al.}(1999)\citenamefont
  {DeMarco}, \citenamefont {Bohn}, \citenamefont {Burke}, \citenamefont
  {Holland},\ and\ \citenamefont {Jin}}]{demarco1999feshbach}%
  \BibitemOpen
  \bibfield  {author} {\bibinfo {author} {\bibfnamefont {B.}~\bibnamefont
  {DeMarco}}, \bibinfo {author} {\bibfnamefont {J.~L.}\ \bibnamefont {Bohn}},
  \bibinfo {author} {\bibfnamefont {J.~P.}\ \bibnamefont {Burke}}, \bibinfo
  {author} {\bibfnamefont {M.}~\bibnamefont {Holland}},\ and\ \bibinfo {author}
  {\bibfnamefont {D.~S.}\ \bibnamefont {Jin}},\ }\bibfield  {title} {\bibinfo
  {title} {Measurement of $\mathit{p}$-wave threshold law using evaporatively
  cooled fermionic atoms},\ }\href
  {https://doi.org/10.1103/PhysRevLett.82.4208} {\bibfield  {journal} {\bibinfo
   {journal} {Phys. Rev. Lett.}\ }\textbf {\bibinfo {volume} {82}},\ \bibinfo
  {pages} {4208} (\bibinfo {year} {1999})}\BibitemShut {NoStop}%
\bibitem [{\citenamefont {Regal}\ \emph {et~al.}(2003)\citenamefont {Regal},
  \citenamefont {Ticknor}, \citenamefont {Bohn},\ and\ \citenamefont
  {Jin}}]{regal2003feshbach}%
  \BibitemOpen
  \bibfield  {author} {\bibinfo {author} {\bibfnamefont {C.~A.}\ \bibnamefont
  {Regal}}, \bibinfo {author} {\bibfnamefont {C.}~\bibnamefont {Ticknor}},
  \bibinfo {author} {\bibfnamefont {J.~L.}\ \bibnamefont {Bohn}},\ and\
  \bibinfo {author} {\bibfnamefont {D.~S.}\ \bibnamefont {Jin}},\ }\bibfield
  {title} {\bibinfo {title} {Tuning $p$-wave interactions in an ultracold
  {F}ermi gas of atoms},\ }\href
  {https://doi.org/10.1103/PhysRevLett.90.053201} {\bibfield  {journal}
  {\bibinfo  {journal} {Phys. Rev. Lett.}\ }\textbf {\bibinfo {volume} {90}},\
  \bibinfo {pages} {053201} (\bibinfo {year} {2003})}\BibitemShut {NoStop}%
\bibitem [{\citenamefont {Luciuk}\ \emph {et~al.}(2016)\citenamefont {Luciuk},
  \citenamefont {Trotzky}, \citenamefont {Smale}, \citenamefont {Yu},
  \citenamefont {Zhang},\ and\ \citenamefont {Thywissen}}]{luciuk2016feshbach}%
  \BibitemOpen
  \bibfield  {author} {\bibinfo {author} {\bibfnamefont {C.}~\bibnamefont
  {Luciuk}}, \bibinfo {author} {\bibfnamefont {S.}~\bibnamefont {Trotzky}},
  \bibinfo {author} {\bibfnamefont {S.}~\bibnamefont {Smale}}, \bibinfo
  {author} {\bibfnamefont {Z.}~\bibnamefont {Yu}}, \bibinfo {author}
  {\bibfnamefont {S.}~\bibnamefont {Zhang}},\ and\ \bibinfo {author}
  {\bibfnamefont {J.~H.}\ \bibnamefont {Thywissen}},\ }\bibfield  {title}
  {\bibinfo {title} {Evidence for universal relations describing a gas with
  p-wave interactions},\ }\href@noop {} {\bibfield  {journal} {\bibinfo
  {journal} {Nat. Phys.}\ }\textbf {\bibinfo {volume} {12}},\ \bibinfo {pages}
  {599} (\bibinfo {year} {2016})}\BibitemShut {NoStop}%
\bibitem [{\citenamefont {Spielman}\ \emph {et~al.}(2006)\citenamefont
  {Spielman}, \citenamefont {Johnson}, \citenamefont {Huckans}, \citenamefont
  {Fertig}, \citenamefont {Rolston}, \citenamefont {Phillips},\ and\
  \citenamefont {Porto}}]{spielman2006collision}%
  \BibitemOpen
  \bibfield  {author} {\bibinfo {author} {\bibfnamefont {I.~B.}\ \bibnamefont
  {Spielman}}, \bibinfo {author} {\bibfnamefont {P.~R.}\ \bibnamefont
  {Johnson}}, \bibinfo {author} {\bibfnamefont {J.~H.}\ \bibnamefont
  {Huckans}}, \bibinfo {author} {\bibfnamefont {C.~D.}\ \bibnamefont {Fertig}},
  \bibinfo {author} {\bibfnamefont {S.~L.}\ \bibnamefont {Rolston}}, \bibinfo
  {author} {\bibfnamefont {W.~D.}\ \bibnamefont {Phillips}},\ and\ \bibinfo
  {author} {\bibfnamefont {J.~V.}\ \bibnamefont {Porto}},\ }\bibfield  {title}
  {\bibinfo {title} {Collisional deexcitation in a quasi-two-dimensional
  degenerate bosonic gas},\ }\href {https://doi.org/10.1103/PhysRevA.73.020702}
  {\bibfield  {journal} {\bibinfo  {journal} {Phys. Rev. A}\ }\textbf {\bibinfo
  {volume} {73}},\ \bibinfo {pages} {020702} (\bibinfo {year}
  {2006})}\BibitemShut {NoStop}%
\bibitem [{\citenamefont {M\"uller}\ \emph {et~al.}(2007)\citenamefont
  {M\"uller}, \citenamefont {F\"olling}, \citenamefont {Widera},\ and\
  \citenamefont {Bloch}}]{muller2007orbital}%
  \BibitemOpen
  \bibfield  {author} {\bibinfo {author} {\bibfnamefont {T.}~\bibnamefont
  {M\"uller}}, \bibinfo {author} {\bibfnamefont {S.}~\bibnamefont {F\"olling}},
  \bibinfo {author} {\bibfnamefont {A.}~\bibnamefont {Widera}},\ and\ \bibinfo
  {author} {\bibfnamefont {I.}~\bibnamefont {Bloch}},\ }\bibfield  {title}
  {\bibinfo {title} {State preparation and dynamics of ultracold atoms in
  higher lattice orbitals},\ }\href
  {https://doi.org/10.1103/PhysRevLett.99.200405} {\bibfield  {journal}
  {\bibinfo  {journal} {Phys. Rev. Lett.}\ }\textbf {\bibinfo {volume} {99}},\
  \bibinfo {pages} {200405} (\bibinfo {year} {2007})}\BibitemShut {NoStop}%
\bibitem [{\citenamefont {Wirth}\ \emph {et~al.}(2011)\citenamefont {Wirth},
  \citenamefont {{\"O}lschl{\"a}ger},\ and\ \citenamefont
  {Hemmerich}}]{wirth2011orbitalSuperfluidity}%
  \BibitemOpen
  \bibfield  {author} {\bibinfo {author} {\bibfnamefont {G.}~\bibnamefont
  {Wirth}}, \bibinfo {author} {\bibfnamefont {M.}~\bibnamefont
  {{\"O}lschl{\"a}ger}},\ and\ \bibinfo {author} {\bibfnamefont
  {A.}~\bibnamefont {Hemmerich}},\ }\bibfield  {title} {\bibinfo {title}
  {Evidence for orbital superfluidity in the p-band of a bipartite optical
  square lattice},\ }\href@noop {} {\bibfield  {journal} {\bibinfo  {journal}
  {Nat. Phys.}\ }\textbf {\bibinfo {volume} {7}},\ \bibinfo {pages} {147}
  (\bibinfo {year} {2011})}\BibitemShut {NoStop}%
\bibitem [{\citenamefont {Kock}\ \emph {et~al.}(2015)\citenamefont {Kock},
  \citenamefont {\"Olschl\"ager}, \citenamefont {Ewerbeck}, \citenamefont
  {Huang}, \citenamefont {Mathey},\ and\ \citenamefont
  {Hemmerich}}]{kock2015orbitalSuperfluidity}%
  \BibitemOpen
  \bibfield  {author} {\bibinfo {author} {\bibfnamefont {T.}~\bibnamefont
  {Kock}}, \bibinfo {author} {\bibfnamefont {M.}~\bibnamefont
  {\"Olschl\"ager}}, \bibinfo {author} {\bibfnamefont {A.}~\bibnamefont
  {Ewerbeck}}, \bibinfo {author} {\bibfnamefont {W.-M.}\ \bibnamefont {Huang}},
  \bibinfo {author} {\bibfnamefont {L.}~\bibnamefont {Mathey}},\ and\ \bibinfo
  {author} {\bibfnamefont {A.}~\bibnamefont {Hemmerich}},\ }\bibfield  {title}
  {\bibinfo {title} {Observing chiral superfluid order by matter-wave
  interference},\ }\href {https://doi.org/10.1103/PhysRevLett.114.115301}
  {\bibfield  {journal} {\bibinfo  {journal} {Phys. Rev. Lett.}\ }\textbf
  {\bibinfo {volume} {114}},\ \bibinfo {pages} {115301} (\bibinfo {year}
  {2015})}\BibitemShut {NoStop}%
\bibitem [{\citenamefont {Di~Liberto}\ \emph {et~al.}(2016)\citenamefont
  {Di~Liberto}, \citenamefont {Hemmerich},\ and\ \citenamefont
  {Morais~Smith}}]{diliberto2016orbitalSuperfluidity}%
  \BibitemOpen
  \bibfield  {author} {\bibinfo {author} {\bibfnamefont {M.}~\bibnamefont
  {Di~Liberto}}, \bibinfo {author} {\bibfnamefont {A.}~\bibnamefont
  {Hemmerich}},\ and\ \bibinfo {author} {\bibfnamefont {C.}~\bibnamefont
  {Morais~Smith}},\ }\bibfield  {title} {\bibinfo {title} {{Topological Varma
  Superfluid in Optical Lattices}},\ }\href
  {https://doi.org/10.1103/PhysRevLett.117.163001} {\bibfield  {journal}
  {\bibinfo  {journal} {Phys. Rev. Lett.}\ }\textbf {\bibinfo {volume} {117}},\
  \bibinfo {pages} {163001} (\bibinfo {year} {2016})}\BibitemShut {NoStop}%
\bibitem [{\citenamefont {Hartke}\ \emph {et~al.}(2021)\citenamefont {Hartke},
  \citenamefont {Oreg}, \citenamefont {Jia},\ and\ \citenamefont
  {Zwierlein}}]{hartke2021dfsFermions}%
  \BibitemOpen
  \bibfield  {author} {\bibinfo {author} {\bibfnamefont {T.}~\bibnamefont
  {Hartke}}, \bibinfo {author} {\bibfnamefont {B.}~\bibnamefont {Oreg}},
  \bibinfo {author} {\bibfnamefont {N.}~\bibnamefont {Jia}},\ and\ \bibinfo
  {author} {\bibfnamefont {M.}~\bibnamefont {Zwierlein}},\ }\bibfield  {title}
  {\bibinfo {title} {Quantum register of fermion pairs},\ }\href@noop {}
  {\bibfield  {journal} {\bibinfo  {journal} {arXiv preprint arXiv:2103.13992}\
  } (\bibinfo {year} {2021})}\BibitemShut {NoStop}%
\bibitem [{\citenamefont {Rey}\ \emph {et~al.}(2008)\citenamefont {Rey},
  \citenamefont {Jiang}, \citenamefont {Fleischhauer}, \citenamefont {Demler},\
  and\ \citenamefont {Lukin}}]{rey2008gap}%
  \BibitemOpen
  \bibfield  {author} {\bibinfo {author} {\bibfnamefont {A.}~\bibnamefont
  {Rey}}, \bibinfo {author} {\bibfnamefont {L.}~\bibnamefont {Jiang}}, \bibinfo
  {author} {\bibfnamefont {M.}~\bibnamefont {Fleischhauer}}, \bibinfo {author}
  {\bibfnamefont {E.}~\bibnamefont {Demler}},\ and\ \bibinfo {author}
  {\bibfnamefont {M.}~\bibnamefont {Lukin}},\ }\bibfield  {title} {\bibinfo
  {title} {Many-body protected entanglement generation in interacting spin
  systems},\ }\href@noop {} {\bibfield  {journal} {\bibinfo  {journal} {Phys.
  Rev. A}\ }\textbf {\bibinfo {volume} {77}},\ \bibinfo {pages} {052305}
  (\bibinfo {year} {2008})}\BibitemShut {NoStop}%
\bibitem [{\citenamefont {Pricoupenko}(2006)}]{Pricoupenko:2006cj}%
  \BibitemOpen
  \bibfield  {author} {\bibinfo {author} {\bibfnamefont {L.}~\bibnamefont
  {Pricoupenko}},\ }\bibfield  {title} {\bibinfo {title} {{Modeling
  interactions for resonant p-wave scattering}},\ }\href
  {https://doi.org/10.1103/PhysRevLett.96.050401} {\bibfield  {journal}
  {\bibinfo  {journal} {Phys. Rev. Lett.}\ }\textbf {\bibinfo {volume} {96}},\
  \bibinfo {pages} {050401} (\bibinfo {year} {2006})}\BibitemShut {NoStop}%
\bibitem [{\citenamefont {Idziaszek}\ and\ \citenamefont
  {Calarco}(2006)}]{Idziaszek:2006gz}%
  \BibitemOpen
  \bibfield  {author} {\bibinfo {author} {\bibfnamefont {Z.}~\bibnamefont
  {Idziaszek}}\ and\ \bibinfo {author} {\bibfnamefont {T.}~\bibnamefont
  {Calarco}},\ }\bibfield  {title} {\bibinfo {title} {{Pseudopotential method
  for higher partial wave scattering}},\ }\href
  {https://doi.org/10.1103/PhysRevLett.96.013201} {\bibfield  {journal}
  {\bibinfo  {journal} {Phys. Rev. Lett.}\ }\textbf {\bibinfo {volume} {96}},\
  \bibinfo {pages} {013201} (\bibinfo {year} {2006})}\BibitemShut {NoStop}%
\bibitem [{\citenamefont {Idziaszek}(2009)}]{Idziaszek:2009jq}%
  \BibitemOpen
  \bibfield  {author} {\bibinfo {author} {\bibfnamefont {Z.}~\bibnamefont
  {Idziaszek}},\ }\bibfield  {title} {\bibinfo {title} {{Analytical solutions
  for two atoms in a harmonic trap: P-wave interactions}},\ }\href
  {https://doi.org/10.1103/PhysRevA.79.062701} {\bibfield  {journal} {\bibinfo
  {journal} {Phys. Rev. A}\ }\textbf {\bibinfo {volume} {79}},\ \bibinfo
  {pages} {062701} (\bibinfo {year} {2009})}\BibitemShut {NoStop}%
\bibitem [{\citenamefont {Zinner}(2012)}]{Zinner:2012gi}%
  \BibitemOpen
  \bibfield  {author} {\bibinfo {author} {\bibfnamefont {N.~T.}\ \bibnamefont
  {Zinner}},\ }\bibfield  {title} {\bibinfo {title} {{Universal two-body
  spectra of ultracold harmonically trapped atoms in two and three
  dimensions}},\ }\href {https://doi.org/10.1088/1751-8113/45/20/205302}
  {\bibfield  {journal} {\bibinfo  {journal} {Journal of Physics A:
  Mathematical and Theoretical}\ }\textbf {\bibinfo {volume} {45}},\ \bibinfo
  {pages} {205302} (\bibinfo {year} {2012})}\BibitemShut {NoStop}%
\bibitem [{SM()}]{SM}%
  \BibitemOpen
  \href@noop {} {}\bibinfo {note} {See Supplemental Material at [hyperlink to
  be provided] for Notes A through F and Figures S1 through S6, including
  References [32,35-36].}\BibitemShut {Stop}%
\bibitem [{\citenamefont {Chiu}\ \emph {et~al.}(2018)\citenamefont {Chiu},
  \citenamefont {Ji}, \citenamefont {Mazurenko}, \citenamefont {Greif},\ and\
  \citenamefont {Greiner}}]{Chiu:2018fl}%
  \BibitemOpen
  \bibfield  {author} {\bibinfo {author} {\bibfnamefont {C.~S.}\ \bibnamefont
  {Chiu}}, \bibinfo {author} {\bibfnamefont {G.}~\bibnamefont {Ji}}, \bibinfo
  {author} {\bibfnamefont {A.}~\bibnamefont {Mazurenko}}, \bibinfo {author}
  {\bibfnamefont {D.}~\bibnamefont {Greif}},\ and\ \bibinfo {author}
  {\bibfnamefont {M.}~\bibnamefont {Greiner}},\ }\bibfield  {title} {\bibinfo
  {title} {{Quantum State Engineering of a {H}ubbard System with Ultracold
  Fermions}},\ }\href {https://doi.org/10.1103/PhysRevLett.120.243201}
  {\bibfield  {journal} {\bibinfo  {journal} {Phys. Rev. Lett.}\ }\textbf
  {\bibinfo {volume} {120}},\ \bibinfo {pages} {243201} (\bibinfo {year}
  {2018})}\BibitemShut {NoStop}%
\bibitem [{\citenamefont {Swallows}\ \emph {et~al.}(2011)\citenamefont
  {Swallows}, \citenamefont {Bishof}, \citenamefont {Lin}, \citenamefont
  {Blatt}, \citenamefont {Martin}, \citenamefont {Rey},\ and\ \citenamefont
  {Ye}}]{swallows2011spinwaveMatrixElements}%
  \BibitemOpen
  \bibfield  {author} {\bibinfo {author} {\bibfnamefont {M.~D.}\ \bibnamefont
  {Swallows}}, \bibinfo {author} {\bibfnamefont {M.}~\bibnamefont {Bishof}},
  \bibinfo {author} {\bibfnamefont {Y.}~\bibnamefont {Lin}}, \bibinfo {author}
  {\bibfnamefont {S.}~\bibnamefont {Blatt}}, \bibinfo {author} {\bibfnamefont
  {M.~J.}\ \bibnamefont {Martin}}, \bibinfo {author} {\bibfnamefont {A.~M.}\
  \bibnamefont {Rey}},\ and\ \bibinfo {author} {\bibfnamefont {J.}~\bibnamefont
  {Ye}},\ }\bibfield  {title} {\bibinfo {title} {Suppression of collisional
  shifts in a strongly interacting lattice clock},\ }\href@noop {} {\bibfield
  {journal} {\bibinfo  {journal} {Science}\ }\textbf {\bibinfo {volume}
  {331}},\ \bibinfo {pages} {1043} (\bibinfo {year} {2011})}\BibitemShut
  {NoStop}%
\bibitem [{\citenamefont {Baier}\ \emph {et~al.}(2016)\citenamefont {Baier},
  \citenamefont {Mark}, \citenamefont {Petter}, \citenamefont {Aikawa},
  \citenamefont {Chomaz}, \citenamefont {Cai}, \citenamefont {Baranov},
  \citenamefont {Zoller},\ and\ \citenamefont {Ferlaino}}]{baier2016extended}%
  \BibitemOpen
  \bibfield  {author} {\bibinfo {author} {\bibfnamefont {S.}~\bibnamefont
  {Baier}}, \bibinfo {author} {\bibfnamefont {M.~J.}\ \bibnamefont {Mark}},
  \bibinfo {author} {\bibfnamefont {D.}~\bibnamefont {Petter}}, \bibinfo
  {author} {\bibfnamefont {K.}~\bibnamefont {Aikawa}}, \bibinfo {author}
  {\bibfnamefont {L.}~\bibnamefont {Chomaz}}, \bibinfo {author} {\bibfnamefont
  {Z.}~\bibnamefont {Cai}}, \bibinfo {author} {\bibfnamefont {M.}~\bibnamefont
  {Baranov}}, \bibinfo {author} {\bibfnamefont {P.}~\bibnamefont {Zoller}},\
  and\ \bibinfo {author} {\bibfnamefont {F.}~\bibnamefont {Ferlaino}},\
  }\bibfield  {title} {\bibinfo {title} {{Extended Bose-Hubbard models with
  ultracold magnetic atoms}},\ }\href@noop {} {\bibfield  {journal} {\bibinfo
  {journal} {Science}\ }\textbf {\bibinfo {volume} {352}},\ \bibinfo {pages}
  {201} (\bibinfo {year} {2016})}\BibitemShut {NoStop}%
\bibitem [{\citenamefont {Lhuillier}\ and\ \citenamefont
  {Laloë}(1982)}]{Laloe}%
  \BibitemOpen
  \bibfield  {author} {\bibinfo {author} {\bibfnamefont {C.}~\bibnamefont
  {Lhuillier}}\ and\ \bibinfo {author} {\bibfnamefont {F.}~\bibnamefont
  {Laloë}},\ }\bibfield  {title} {\bibinfo {title} {Transport properties in a
  spin polarized gas, i},\ }\href
  {https://doi.org/10.1051/jphys:01982004302019700} {\bibfield  {journal}
  {\bibinfo  {journal} {J. Phys. France}\ }\textbf {\bibinfo {volume} {43}},\
  \bibinfo {pages} {197} (\bibinfo {year} {1982})}\BibitemShut {NoStop}%
\bibitem [{\citenamefont {Rey}\ \emph {et~al.}(2014)\citenamefont {Rey},
  \citenamefont {Gorshkov}, \citenamefont {Kraus}, \citenamefont {Martin},
  \citenamefont {Bishof}, \citenamefont {Swallows}, \citenamefont {Zhang},
  \citenamefont {Benko}, \citenamefont {Ye}, \citenamefont {Lemke},\ and\
  \citenamefont {Ludlow}}]{Rey2014}%
  \BibitemOpen
  \bibfield  {author} {\bibinfo {author} {\bibfnamefont {A.}~\bibnamefont
  {Rey}}, \bibinfo {author} {\bibfnamefont {A.}~\bibnamefont {Gorshkov}},
  \bibinfo {author} {\bibfnamefont {C.}~\bibnamefont {Kraus}}, \bibinfo
  {author} {\bibfnamefont {M.}~\bibnamefont {Martin}}, \bibinfo {author}
  {\bibfnamefont {M.}~\bibnamefont {Bishof}}, \bibinfo {author} {\bibfnamefont
  {M.}~\bibnamefont {Swallows}}, \bibinfo {author} {\bibfnamefont
  {X.}~\bibnamefont {Zhang}}, \bibinfo {author} {\bibfnamefont
  {C.}~\bibnamefont {Benko}}, \bibinfo {author} {\bibfnamefont
  {J.}~\bibnamefont {Ye}}, \bibinfo {author} {\bibfnamefont {N.}~\bibnamefont
  {Lemke}},\ and\ \bibinfo {author} {\bibfnamefont {A.}~\bibnamefont
  {Ludlow}},\ }\bibfield  {title} {\bibinfo {title} {Probing many-body
  interactions in an optical lattice clock},\ }\href
  {https://doi.org/https://doi.org/10.1016/j.aop.2013.11.002} {\bibfield
  {journal} {\bibinfo  {journal} {Annals of Physics}\ }\textbf {\bibinfo
  {volume} {340}},\ \bibinfo {pages} {311} (\bibinfo {year}
  {2014})}\BibitemShut {NoStop}%
\bibitem [{\citenamefont {K{\"o}hl}\ \emph {et~al.}(2005)\citenamefont
  {K{\"o}hl}, \citenamefont {Moritz}, \citenamefont {St{\"o}ferle},
  \citenamefont {G{\"u}nter},\ and\ \citenamefont
  {Esslinger}}]{kohl2005bandMapping}%
  \BibitemOpen
  \bibfield  {author} {\bibinfo {author} {\bibfnamefont {M.}~\bibnamefont
  {K{\"o}hl}}, \bibinfo {author} {\bibfnamefont {H.}~\bibnamefont {Moritz}},
  \bibinfo {author} {\bibfnamefont {T.}~\bibnamefont {St{\"o}ferle}}, \bibinfo
  {author} {\bibfnamefont {K.}~\bibnamefont {G{\"u}nter}},\ and\ \bibinfo
  {author} {\bibfnamefont {T.}~\bibnamefont {Esslinger}},\ }\bibfield  {title}
  {\bibinfo {title} {Fermionic atoms in a three dimensional optical lattice:
  Observing fermi surfaces, dynamics, and interactions},\ }\href@noop {}
  {\bibfield  {journal} {\bibinfo  {journal} {Phys. Rev. Lett.}\ }\textbf
  {\bibinfo {volume} {94}},\ \bibinfo {pages} {080403} (\bibinfo {year}
  {2005})}\BibitemShut {NoStop}%
\bibitem [{\citenamefont {Smale}\ \emph {et~al.}(2019)\citenamefont {Smale},
  \citenamefont {He}, \citenamefont {Olsen}, \citenamefont {Jackson},
  \citenamefont {Sharum}, \citenamefont {Trotzky}, \citenamefont {Marino},
  \citenamefont {Rey},\ and\ \citenamefont
  {Thywissen}}]{smale2019gapProtectionDPT}%
  \BibitemOpen
  \bibfield  {author} {\bibinfo {author} {\bibfnamefont {S.}~\bibnamefont
  {Smale}}, \bibinfo {author} {\bibfnamefont {P.}~\bibnamefont {He}}, \bibinfo
  {author} {\bibfnamefont {B.~A.}\ \bibnamefont {Olsen}}, \bibinfo {author}
  {\bibfnamefont {K.~G.}\ \bibnamefont {Jackson}}, \bibinfo {author}
  {\bibfnamefont {H.}~\bibnamefont {Sharum}}, \bibinfo {author} {\bibfnamefont
  {S.}~\bibnamefont {Trotzky}}, \bibinfo {author} {\bibfnamefont
  {J.}~\bibnamefont {Marino}}, \bibinfo {author} {\bibfnamefont {A.~M.}\
  \bibnamefont {Rey}},\ and\ \bibinfo {author} {\bibfnamefont {J.~H.}\
  \bibnamefont {Thywissen}},\ }\bibfield  {title} {\bibinfo {title}
  {{Observation of a transition between dynamical phases in a quantum
  degenerate Fermi gas}},\ }\href@noop {} {\bibfield  {journal} {\bibinfo
  {journal} {Science Adv.}\ }\textbf {\bibinfo {volume} {5}},\ \bibinfo {pages}
  {eaax1568} (\bibinfo {year} {2019})}\BibitemShut {NoStop}%
\bibitem [{\citenamefont {Chu}\ \emph {et~al.}(2020)\citenamefont {Chu},
  \citenamefont {Will}, \citenamefont {Arlt}, \citenamefont {Klempt},\ and\
  \citenamefont {Rey}}]{chu2020gap}%
  \BibitemOpen
  \bibfield  {author} {\bibinfo {author} {\bibfnamefont {A.}~\bibnamefont
  {Chu}}, \bibinfo {author} {\bibfnamefont {J.}~\bibnamefont {Will}}, \bibinfo
  {author} {\bibfnamefont {J.}~\bibnamefont {Arlt}}, \bibinfo {author}
  {\bibfnamefont {C.}~\bibnamefont {Klempt}},\ and\ \bibinfo {author}
  {\bibfnamefont {A.~M.}\ \bibnamefont {Rey}},\ }\bibfield  {title} {\bibinfo
  {title} {Simulation of {XXZ} spin models using sideband transitions in
  trapped bosonic gases},\ }\href@noop {} {\bibfield  {journal} {\bibinfo
  {journal} {Phys. Rev. Lett.}\ }\textbf {\bibinfo {volume} {125}},\ \bibinfo
  {pages} {240504} (\bibinfo {year} {2020})}\BibitemShut {NoStop}%
\bibitem [{\citenamefont {Kitagawa}\ and\ \citenamefont
  {Ueda}(1993)}]{kitagawa1993squeezed}%
  \BibitemOpen
  \bibfield  {author} {\bibinfo {author} {\bibfnamefont {M.}~\bibnamefont
  {Kitagawa}}\ and\ \bibinfo {author} {\bibfnamefont {M.}~\bibnamefont
  {Ueda}},\ }\bibfield  {title} {\bibinfo {title} {Squeezed spin states},\
  }\href@noop {} {\bibfield  {journal} {\bibinfo  {journal} {Phys. Rev. A}\
  }\textbf {\bibinfo {volume} {47}},\ \bibinfo {pages} {5138} (\bibinfo {year}
  {1993})}\BibitemShut {NoStop}%
\bibitem [{\citenamefont {Dutta}\ \emph {et~al.}(2015)\citenamefont {Dutta},
  \citenamefont {Gajda}, \citenamefont {Hauke}, \citenamefont {Lewenstein},
  \citenamefont {L{\"u}hmann}, \citenamefont {Malomed}, \citenamefont
  {Sowi{\'n}ski},\ and\ \citenamefont {Zakrzewski}}]{dutta2015extendedHubbard}%
  \BibitemOpen
  \bibfield  {author} {\bibinfo {author} {\bibfnamefont {O.}~\bibnamefont
  {Dutta}}, \bibinfo {author} {\bibfnamefont {M.}~\bibnamefont {Gajda}},
  \bibinfo {author} {\bibfnamefont {P.}~\bibnamefont {Hauke}}, \bibinfo
  {author} {\bibfnamefont {M.}~\bibnamefont {Lewenstein}}, \bibinfo {author}
  {\bibfnamefont {D.-S.}\ \bibnamefont {L{\"u}hmann}}, \bibinfo {author}
  {\bibfnamefont {B.~A.}\ \bibnamefont {Malomed}}, \bibinfo {author}
  {\bibfnamefont {T.}~\bibnamefont {Sowi{\'n}ski}},\ and\ \bibinfo {author}
  {\bibfnamefont {J.}~\bibnamefont {Zakrzewski}},\ }\bibfield  {title}
  {\bibinfo {title} {Non-standard {H}ubbard models in optical lattices: a
  review},\ }\href@noop {} {\bibfield  {journal} {\bibinfo  {journal} {Rep.
  Prog. Phys.}\ }\textbf {\bibinfo {volume} {78}},\ \bibinfo {pages} {066001}
  (\bibinfo {year} {2015})}\BibitemShut {NoStop}%
\bibitem [{\citenamefont {Parsons}\ \emph {et~al.}(2015)\citenamefont
  {Parsons}, \citenamefont {Huber}, \citenamefont {Mazurenko}, \citenamefont
  {Chiu}, \citenamefont {Setiawan}, \citenamefont {Wooley-Brown}, \citenamefont
  {Blatt},\ and\ \citenamefont {Greiner}}]{parsons2015microscope}%
  \BibitemOpen
  \bibfield  {author} {\bibinfo {author} {\bibfnamefont {M.~F.}\ \bibnamefont
  {Parsons}}, \bibinfo {author} {\bibfnamefont {F.}~\bibnamefont {Huber}},
  \bibinfo {author} {\bibfnamefont {A.}~\bibnamefont {Mazurenko}}, \bibinfo
  {author} {\bibfnamefont {C.~S.}\ \bibnamefont {Chiu}}, \bibinfo {author}
  {\bibfnamefont {W.}~\bibnamefont {Setiawan}}, \bibinfo {author}
  {\bibfnamefont {K.}~\bibnamefont {Wooley-Brown}}, \bibinfo {author}
  {\bibfnamefont {S.}~\bibnamefont {Blatt}},\ and\ \bibinfo {author}
  {\bibfnamefont {M.}~\bibnamefont {Greiner}},\ }\bibfield  {title} {\bibinfo
  {title} {{Site-resolved imaging of fermionic Li-6 in an optical lattice}},\
  }\href@noop {} {\bibfield  {journal} {\bibinfo  {journal} {Phys. Rev. Lett.}\
  }\textbf {\bibinfo {volume} {114}},\ \bibinfo {pages} {213002} (\bibinfo
  {year} {2015})}\BibitemShut {NoStop}%
\bibitem [{\citenamefont {Haller}\ \emph {et~al.}(2015)\citenamefont {Haller},
  \citenamefont {Hudson}, \citenamefont {Kelly}, \citenamefont {Cotta},
  \citenamefont {Peaudecerf}, \citenamefont {Bruce},\ and\ \citenamefont
  {Kuhr}}]{haller2015microscope}%
  \BibitemOpen
  \bibfield  {author} {\bibinfo {author} {\bibfnamefont {E.}~\bibnamefont
  {Haller}}, \bibinfo {author} {\bibfnamefont {J.}~\bibnamefont {Hudson}},
  \bibinfo {author} {\bibfnamefont {A.}~\bibnamefont {Kelly}}, \bibinfo
  {author} {\bibfnamefont {D.~A.}\ \bibnamefont {Cotta}}, \bibinfo {author}
  {\bibfnamefont {B.}~\bibnamefont {Peaudecerf}}, \bibinfo {author}
  {\bibfnamefont {G.~D.}\ \bibnamefont {Bruce}},\ and\ \bibinfo {author}
  {\bibfnamefont {S.}~\bibnamefont {Kuhr}},\ }\bibfield  {title} {\bibinfo
  {title} {Single-atom imaging of fermions in a quantum-gas microscope},\
  }\href@noop {} {\bibfield  {journal} {\bibinfo  {journal} {Nat. Phys.}\
  }\textbf {\bibinfo {volume} {11}},\ \bibinfo {pages} {738} (\bibinfo {year}
  {2015})}\BibitemShut {NoStop}%
\bibitem [{\citenamefont {Cheuk}\ \emph {et~al.}(2015)\citenamefont {Cheuk},
  \citenamefont {Nichols}, \citenamefont {Okan}, \citenamefont {Gersdorf},
  \citenamefont {Ramasesh}, \citenamefont {Bakr}, \citenamefont {Lompe},\ and\
  \citenamefont {Zwierlein}}]{cheuk2015microscope}%
  \BibitemOpen
  \bibfield  {author} {\bibinfo {author} {\bibfnamefont {L.~W.}\ \bibnamefont
  {Cheuk}}, \bibinfo {author} {\bibfnamefont {M.~A.}\ \bibnamefont {Nichols}},
  \bibinfo {author} {\bibfnamefont {M.}~\bibnamefont {Okan}}, \bibinfo {author}
  {\bibfnamefont {T.}~\bibnamefont {Gersdorf}}, \bibinfo {author}
  {\bibfnamefont {V.~V.}\ \bibnamefont {Ramasesh}}, \bibinfo {author}
  {\bibfnamefont {W.~S.}\ \bibnamefont {Bakr}}, \bibinfo {author}
  {\bibfnamefont {T.}~\bibnamefont {Lompe}},\ and\ \bibinfo {author}
  {\bibfnamefont {M.~W.}\ \bibnamefont {Zwierlein}},\ }\bibfield  {title}
  {\bibinfo {title} {Quantum-gas microscope for fermionic atoms},\ }\href@noop
  {} {\bibfield  {journal} {\bibinfo  {journal} {Phys. Rev. Lett.}\ }\textbf
  {\bibinfo {volume} {114}},\ \bibinfo {pages} {193001} (\bibinfo {year}
  {2015})}\BibitemShut {NoStop}%
\bibitem [{\citenamefont {Omran}\ \emph {et~al.}(2015)\citenamefont {Omran},
  \citenamefont {Boll}, \citenamefont {Hilker}, \citenamefont {Kleinlein},
  \citenamefont {Salomon}, \citenamefont {Bloch},\ and\ \citenamefont
  {Gross}}]{omran2015microscope}%
  \BibitemOpen
  \bibfield  {author} {\bibinfo {author} {\bibfnamefont {A.}~\bibnamefont
  {Omran}}, \bibinfo {author} {\bibfnamefont {M.}~\bibnamefont {Boll}},
  \bibinfo {author} {\bibfnamefont {T.~A.}\ \bibnamefont {Hilker}}, \bibinfo
  {author} {\bibfnamefont {K.}~\bibnamefont {Kleinlein}}, \bibinfo {author}
  {\bibfnamefont {G.}~\bibnamefont {Salomon}}, \bibinfo {author} {\bibfnamefont
  {I.}~\bibnamefont {Bloch}},\ and\ \bibinfo {author} {\bibfnamefont
  {C.}~\bibnamefont {Gross}},\ }\bibfield  {title} {\bibinfo {title}
  {{Microscopic observation of Pauli blocking in degenerate fermionic lattice
  gases}},\ }\href@noop {} {\bibfield  {journal} {\bibinfo  {journal} {Phys.
  Rev. Lett.}\ }\textbf {\bibinfo {volume} {115}},\ \bibinfo {pages} {263001}
  (\bibinfo {year} {2015})}\BibitemShut {NoStop}%
\bibitem [{\citenamefont {Edge}\ \emph {et~al.}(2015)\citenamefont {Edge},
  \citenamefont {Anderson}, \citenamefont {Jervis}, \citenamefont {McKay},
  \citenamefont {Day}, \citenamefont {Trotzky},\ and\ \citenamefont
  {Thywissen}}]{edge2015microscope}%
  \BibitemOpen
  \bibfield  {author} {\bibinfo {author} {\bibfnamefont {G.~J.}\ \bibnamefont
  {Edge}}, \bibinfo {author} {\bibfnamefont {R.}~\bibnamefont {Anderson}},
  \bibinfo {author} {\bibfnamefont {D.}~\bibnamefont {Jervis}}, \bibinfo
  {author} {\bibfnamefont {D.~C.}\ \bibnamefont {McKay}}, \bibinfo {author}
  {\bibfnamefont {R.}~\bibnamefont {Day}}, \bibinfo {author} {\bibfnamefont
  {S.}~\bibnamefont {Trotzky}},\ and\ \bibinfo {author} {\bibfnamefont {J.~H.}\
  \bibnamefont {Thywissen}},\ }\bibfield  {title} {\bibinfo {title} {Imaging
  and addressing of individual fermionic atoms in an optical lattice},\
  }\href@noop {} {\bibfield  {journal} {\bibinfo  {journal} {Phys. Rev. A}\
  }\textbf {\bibinfo {volume} {92}},\ \bibinfo {pages} {063406} (\bibinfo
  {year} {2015})}\BibitemShut {NoStop}%
\end{thebibliography}%

\clearpage
\onecolumngrid

\begin{center}
\LARGE
Supplemental Material for "Collective P-Wave Orbital Dynamics of Ultracold Fermions"
\normalsize
\end{center}

\appendixpageoff
\appendixtitleoff
\begin{appendices}
\renewcommand{\thefigure}{S\arabic{figure}}
\setcounter{figure}{0}
\renewcommand \thesubsection {\thesection.\arabic{subsection}}

\section{P-Wave Fermi-Hubbard model derivation}
\label{app_FermiDerivation}
\renewcommand{\theequation}{A\arabic{equation}}
\setcounter{equation}{0}
\subsection{Pseudo-potential and two-body comparison}
In this supplementary we show a comparison between the Wannier basis lattice interactions used in the main text and a more microscopic two-atom picture. We follow Ref.~\cite{Idziaszek:2009jq}, considering the exact solution for two identical fermions in an axially-symmetric 3D harmonic trap (corresponding to a single lattice site). The Hamiltonian describing the two atoms can be written as
\begin{equation}
\begin{aligned}
    \hat{H}_{2\> \mathrm{atom}} =& -\frac{\hbar^2}{2\mu}\vec{\nabla}_{\vec{r}}^2 +\frac{1}{2}\mu \omega^2 \eta^2 \left(r_X^2 + r_Y^2\right) + \frac{1}{2}\mu \omega^2 r_Z^2\\
    &+\frac{\pi \hbar^2}{\mu}b_{XY}^3(E) \vec{\nabla}^{\leftarrow}_{\vec{r}}\delta^{(3)}(\vec{r})\frac{\partial^3}{\partial r^3}r^3, \vec{\nabla}_{\vec{r}}^{\rightarrow}\\
    &+\hat{H}_{\mathrm{COM}}.
\end{aligned}
\end{equation}
Here $\vec{r} = \vec{r}_1 - \vec{r}_2 = (r_X,r_Y,r_Z)$ is the relative position (for atom positions $\vec{r}_1$, $\vec{r}_2$), $\mu = m/2$ the reduced mass, and $(\eta \omega, \eta \omega, \omega)$ the trapping frequencies along the $(X,Y,Z)$ directions (with $X$, $Y$ differing from $Z$ by a dimensionless ratio $\eta$). The second line has the explicit pseudo-potential in the $m_l = \pm 1$ collisional channels (thus proportional to the transverse scattering volume $b_{XY}^3$ in our context), where $r = |\vec{r}|$, $\vec{\nabla}^{\leftarrow}_{\vec{r}}$ and $\vec{\nabla}^{\rightarrow}_{\vec{r}}$ are left- and right-acting gradient operators, and $\delta^{(3)}(\vec{r})$ is the 3D Dirac delta function. The last line contains the center-of-mass contribution $\hat{H}_{\mathrm{COM}} =-\frac{\hbar^2}{8\mu}\vec{\nabla}_{\vec{r}_{c}}^2 +2\mu \omega^2 \eta^2 \left[(\vec{r}_{c})_X^2 + (\vec{r}_{c})_Y^2\right] + 2\mu \omega^2 (\vec{r}_{c})_Z^2$ for $\vec{r}_c = (\vec{r}_1 + \vec{r}_2)/2$, which amounts to a constant shift in energy.

We assume the use of a magnetic Feshbach resonance to tune $b_{XY}^3$, approximating its form as,
\begin{equation}
\label{eq_scatteringVolume}
b_{XY}^3(E) = V_{bg} \left(1 - \frac{\Delta B}{\delta B - \frac{E}{\delta\mu_B}}\right),
\end{equation}
where $V_{bg}$ is the background p-wave scattering volume for $m_{l}=\pm 1$, $\Delta B$ the width of the Feshbach resonance, $\delta B$ the magnetic field detuning away from the resonance, $E$ the energy of the two-atom system $\hat{H}_{2\>\mathrm{atom}}$, and $\delta \mu_B$ the difference in magnetic moment between the open and closed channels. Note that unlike the s-wave case the energy dependence cannot be omitted, as $E/\delta\mu_B$ can be comparable to typical shifts $\delta B$ needed for a non-negligible amplification to the scattering volume. All of the parameters except energy $E$ are experimental atom-dependent values. We use the sample atom of $^{40}$K, which has $V_{bg} = (-107.4 a_0)^3$ with $a_0$ the Bohr radius, $\Delta B = -19.5$ G, $\delta \mu_B = 93$ kHz/G.

\begin{figure}[tb]
\centering
\includegraphics[width=1.0\linewidth]{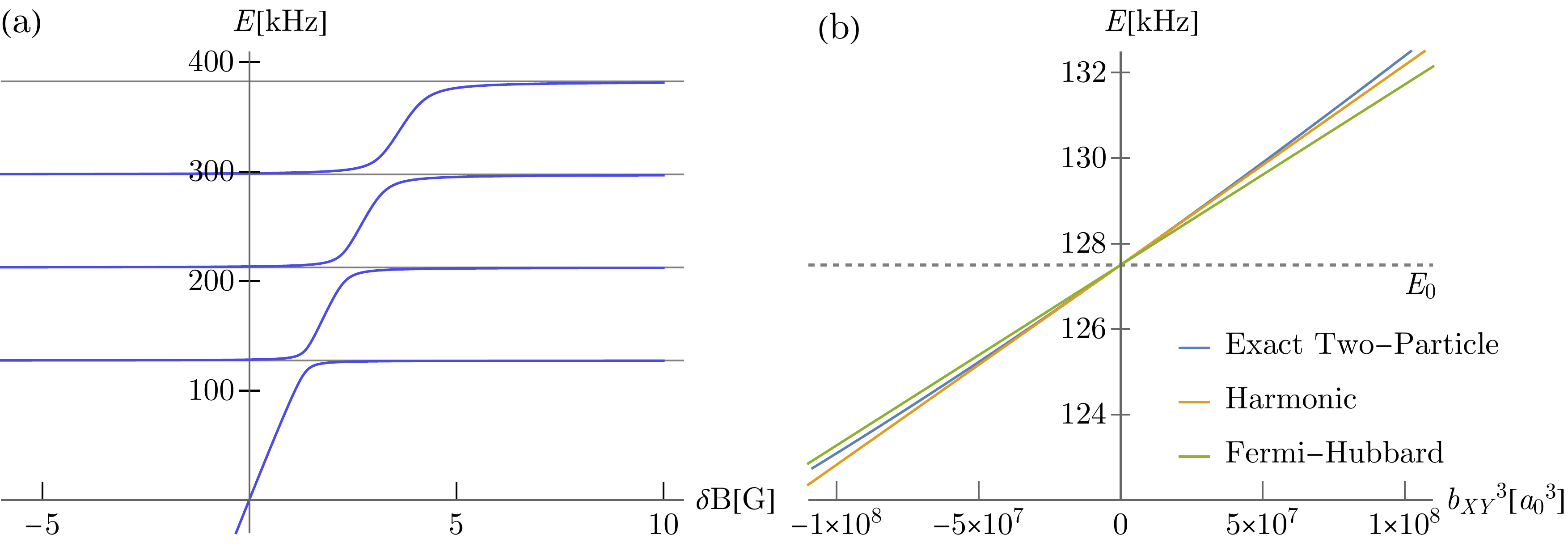}
\caption{(a) Spectrum of two identical fermions in an axially-symmetric harmonic trap with $\omega = 85$ kHz, $\eta = 1/2$. The gray lines correspond to the allowed relative-coordinate energies of $\hat{H}_{2\>\mathrm{atom}}$ in the non-interacting limit $b_{XY}^3 \to 0$. (b) Lowest branch of the spectrum as a function of scattering volume. The blue line is the exact two-atom result obtained by inverting the implicit numerical solutions of the previous panel. The orange line gives the approximation from Eq.~\eqref{eq_FermiHubbardTwoAtomEnergy} using harmonic oscillator eigenfunctions as the on-site orbitals, while the green line gives the same approximation using lattice Wannier functions (as in the main text calculations).}
\label{fig_PseudopotentialComparison}
\end{figure}

Following Ref.~\cite{Idziaszek:2009jq}, the energy of $\hat{H}_{2\>\mathrm{atom}}$ and the scattering volume $b_{XY}^3(E)$ are related by an implicit set of equations that can be solved numerically for the magnetic field detuning $\delta B$ and energy $E$. In Fig.~\ref{fig_PseudopotentialComparison}(a) we plot a typical spectrum for trapping frequencies roughly corresponding to the sample parameters used in the main text ($\omega = 85$ kHz, $\eta = 1/2$). Since we want to operate in the linear regime where the interaction energy shift is small and the wavefunctions of the oscillator are mostly unperturbed, it is also useful to compute the
spectrum as a function of scattering volume $b_{XY}^3$. The blue line in Fig.~\ref{fig_PseudopotentialComparison}(b) shows the lowest branch of the spectrum (corresponding to the first excitation of the relative coordinate allowed by its spatial symmetries) for small scattering volumes.

To connect with the lattice picture, we compare a specific on-site interaction coefficient $U_{\Uparrow\Downarrow}$, which is the on-site Fermi-Hubbard energy of two singly-excited particles. The full two-atom energy is \begin{equation}
\label{eq_FermiHubbardTwoAtomEnergy}
E = E_0 + U_{\Uparrow\Downarrow} \approx E_0 - \frac{3 \pi \hbar^2}{2m} b_{XY}^3(E_0)\>\cdot 4\sum_{\nu = X,Y}\int d^3 \vec{R} W_{\nu}\left[\phi^{\Uparrow *}(\vec{R}),\phi^{\Downarrow *}(\vec{R})\right]W_{\nu}\left[\phi^{\Downarrow *}(\vec{R}),\phi^{\Uparrow}(\vec{R})\right],
\end{equation}
where $E_0= \frac{1}{2}\left(1+4\eta\right)\hbar \omega$ is the un-perturbed oscillator energy (note that it is only half of the two-particle non-interacting energy, with the other half carried by the center-of-mass motion), and the rest of the equation corresponds to the Fermi-Hubbard interaction shift. The factor of 4 in front of the integral comes from the fact that we have four different terms in the Wannier expansion that yield the same contribution. Note that we still have an energy-dependent scattering volume here, but have approximated the energy dependence to be just the bare oscillator energy, since we want to be in the regime where $U_{\Uparrow\Downarrow} \ll E_0$. To leading order, we can account for this energy dependence by shifting the position of the Feshbach resonance (the denominator of Eq.~\eqref{eq_scatteringVolume}) by a corresponding amount $E_0 / \delta \mu_B$. The interaction energy shift is linearly proportional to the scattering volume in this approximation. The orange line in Fig.~\ref{fig_PseudopotentialComparison} shows the approximate Fermi-Hubbard two atom energy using harmonic oscillator wavefunctions with the same trapping frequencies $(\eta \omega, \eta\omega, \omega)$ as the exact caculation in the overlap integral of Eq.~\eqref{eq_FermiHubbardTwoAtomEnergy} (to connect with the exact solution). The green line gives the same approximation using Wannier orbitals in Eq.~\eqref{eq_FermiHubbardTwoAtomEnergy}, matching the calculations used in the main text. We find good agreement for small scattering volumes $|b_{XY}^3| \lesssim 10^8 a_0$, corresponding to $|b_{XY}| \lesssim 400 a_0$.

To connect with our main text results, we also give the interaction shift $U_{\Uparrow\Downarrow}$ for roughly the parameters used in the calculations of the main text, $b_{XY}^3 = (292 a_0)^3= 2.5\times 10^7 a_0^3$ and lattice depths of $V_Z = 100$, $V_X = V_Y = 28$ (for these lattice depths, $\eta \approx 1/2$ and the analytic results of Ref.~\cite{Idziaszek:2009jq} are simpler). With this scattering volume and lattice depth, the exact two-particle calculation using an axial harmonic trap [the blue line in Fig.~\ref{fig_PseudopotentialComparison}(b)] yields an interaction energy of $U_{\Uparrow\Downarrow}\approx 1180$ Hz, the approximate Fermi-Hubbard shift using harmonic oscillator orbitals (orange line) gives $U_{\Uparrow\Downarrow} \approx 1160$ Hz, and the approximate Fermi-Hubbard shift using Wannier functions gives $U_{\Uparrow\Downarrow} \approx 1050$ Hz. Note that in these calculations, for given lattice depths $(V_X, V_Y, V_Z)$ we compute the lattice Wannier functions for using standard methods, then obtain corresponding harmonic trapping frequencies $(\omega_X, \omega_Y, \omega_Z) = (\eta \omega, \eta\omega, \omega)$ (for $V_X = V_Y)$ by matching each direction's frequency to the band gap between the ground and first-excited band. This is not quite the same as conventional approaches which simply Taylor expand the lattice potential $V_{\nu} E_r \sin^2(\pi \nu/a)$ and match it to a harmonic potential. By setting the band gap as the harmonic oscillator frequency, we better capture the anharmonicity of the lattice potential, since we involve higher bands in our calculations and are considering lattice depths that are not quite in the $V_{\nu} \to \infty$ limit.

While the Wannier result is more physically relevant for the lattice system, one can also use the harmonic wavefunction result for rough analytic estimates of the interaction strength. For instance, our shift $U_{\Uparrow\Downarrow}$ using harmonic wavefunctions can be written as,
\begin{equation}
    U_{\Uparrow\Downarrow} \approx \frac{3\sqrt{2} m^{3/2} \eta^2 \omega^{5/2}}{\sqrt{\pi \hbar}} b_{XY}^3.
\end{equation}

\subsection{P-wave interaction coefficients}
Having justified the Fermi-Hubbard approximation for the p-wave interactions, we now show which interaction coefficients $U_{\vec{r},\vec{r}',\vec{r}'',\vec{r}'''}^{\alpha\beta\sigma\gamma}$ in the main text are most relevant to the dynamics.

\begin{figure}[tb]
\centering
\includegraphics[width=1.0\linewidth]{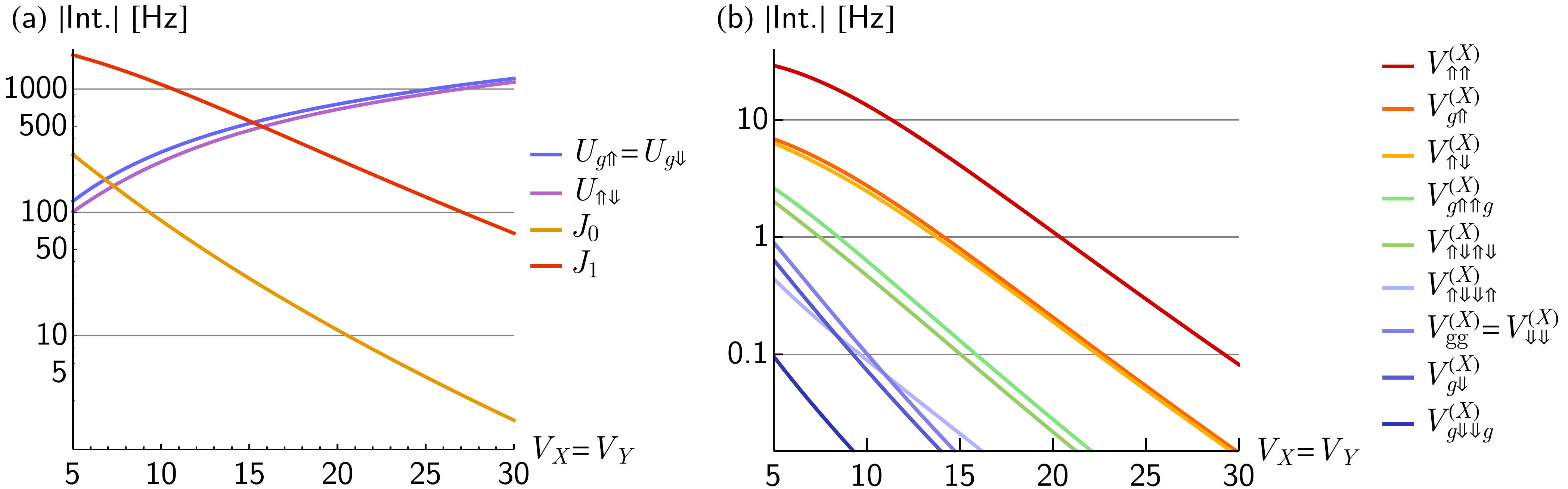}
\caption{(a) On-site interaction parameter and lattice tunneling rate magnitudes as a function of $V_X = V_Y$ lattice depth in units of recoil energy. Transverse lattice depth is fixed at $V_Z = 100$, and scattering length set to $b_{XY} = 292 a_0$. Overlap integrals are computed using ideal Wannier functions. (b) Nearest-neighbour interaction parameters along the $X$-direction. The $V_{\Uparrow\Uparrow}^{(X)}$, $V_{g\Uparrow}^{(X)}$, $V_{\Uparrow\Downarrow}^{(X)}$ terms are strongest, and we will neglect all others.}
\label{fig_InteractionParameters}
\end{figure}

The on-site interaction terms $\hat{H}_{\mathrm{int}}^{\mathrm{(OS)}}$ consist of all terms where $\vec{r} = \vec{r}'  = \vec{r}'' = \vec{r}'''$. All such terms with non-vanishing matrix elements are
\begin{equation}
\label{eq_AppOnSiteInteractions}
\hat{H}_{\mathrm{int}}^{\mathrm{(OS)}}=\sum_{\vec{r}} \left[U_{g\Uparrow}\hat{n}_{\vec{r},g}\hat{n}_{\vec{r},\Uparrow}+U_{g\Downarrow}\hat{n}_{\vec{r},g}\hat{n}_{\vec{r},\Downarrow}+U_{\Uparrow\Downarrow}\hat{n}_{\vec{r},\Uparrow}\hat{n}_{\vec{r},\Downarrow}\right].
\end{equation}
Here $U_{g\Uparrow} = 4 U_{\vec{r},\vec{r},\vec{r},\vec{r}}^{g\Uparrow\Uparrow g}$, $U_{g\Downarrow}=4 U_{\vec{r},\vec{r},\vec{r},\vec{r}}^{g\Downarrow\Downarrow g}$ (with $U_{g\Uparrow} = U_{g\Downarrow}$ because we consider equal lattice depths $V_X = V_Y$) and $U_{\Uparrow\Downarrow}=4 U_{\vec{r},\vec{r},\vec{r},\vec{r}}^{\Uparrow\Downarrow\Downarrow \Uparrow}$, with factors of 4 as before to account for different permutations yielding the same interaction term. Fig.~\ref{fig_InteractionParameters}(a) shows a numerical comparison of these interaction strength magnitudes together with the tunneling rates $J_0$, $J_1$ as a function of the lattice depths $V_X = V_Y$ (fixing $V_Z = 100$ and $b_{XY}=292 a_0$ as in the main text). Naturally, the on-site interaction strengths increase with tighter confinement while the tunneling decreases exponentially.

These interactions can be simplified further since the ground band is completely filled. The $g$ atoms cannot tunnel due to Pauli exclusion, and there are no energy-conserving p-wave interaction processes that can kick them out of their band due to the band gap. We thus assume that all $g$ atoms will sit in their respective lattice sites throughout the time-evolution and make the approximation of,
\begin{equation}
\label{eq_GroundBandFrozen}
    \hat{c}_{\vec{r},g}^{\dagger}\hat{c}_{\vec{r}',g} = \delta_{\vec{r},\vec{r}'} \mathbbm{1}.
\end{equation}
Under this approximation, the $U_{g\Uparrow}$ and $U_{g\Downarrow}$ interaction terms become single-particle shifts $\sum_{\vec{r}}(U_{g\Uparrow}\hat{n}_{\vec{r},\Uparrow}+U_{g\Downarrow}\hat{n}_{\vec{r},\Downarrow})$. Since $U_{g\Uparrow} = U_{g\Downarrow}$, this is proportional to the total number of excited atoms $\sum_{\vec{r}}(\hat{n}_{\vec{r},\Uparrow}+\hat{n}_{\vec{r},\Downarrow})$, which is conserved. We thus drop the $U_{g\Uparrow}$, $U_{g\Downarrow}$ terms altogether, leaving only the $U_{\Uparrow\Downarrow}$-proportional interaction as in the main text.

The next set of terms is cross-site interactions, for which the spatial index $\vec{r}$ is not the same for all four operators. Before writing them down, we first identify several types of terms which are neglected from the start:
\begin{itemize}
\item Any terms that do not conserve the fixed population of two atoms per lattice site, such as paired hops $\hat{c}_{\vec{r},\Uparrow}^{\dagger}\hat{c}_{\vec{r},\Downarrow}^{\dagger}\hat{c}_{\vec{r}+\vec{r}_X,\Uparrow}\hat{c}_{\vec{r} + \vec{r}_X,\Downarrow}$ or interaction-assisted tunneling $\hat{n}_{\vec{r},\Uparrow}(\hat{c}_{\vec{r},\Downarrow}^{\dagger}\hat{c}_{\vec{r}+\vec{r}_X,\Downarrow}+h.c.)$. Such terms are energetically suppressed by the stronger on-site interactions provided that the lattice is not too shallow.
\item Terms that move atoms in/out of the ground band $g$, such as interaction-assisted flips $\hat{n}_{\vec{r},\Uparrow}(\hat{c}_{\vec{r}+\vec{r}_X,\Downarrow}^{\dagger}\hat{c}_{\vec{r}+\vec{r}_X,g}+h.c.)$ or double-flips $\hat{c}_{\vec{r},\Uparrow}^{\dagger}\hat{c}_{\vec{r},g}\hat{c}_{\vec{r}+\vec{r}_X,\Uparrow}^{\dagger}\hat{c}_{\vec{r}+\vec{r}_X,g}$, which are suppressed by the band gap.
\item Any terms that are further range than nearest-neighbour. These are suppressed by exponential falloff of the Wannier functions provided that the lattice is not too shallow.
\end{itemize}

The remaining non-neglected nearest-neighbour interactions can be written as
\footnotesize
\begin{equation}
\label{eq_AppCrossSiteInteractions}
\begin{aligned}
\hat{H}^{\mathrm{NN}}_{\mathrm{int}}&=\normalsize{\sum_{\nu = X,Y}}\bigg[ V_{gg}^{(\nu)}\sum_{\vec{r}}\hat{n}_{\vec{r},g}\hat{n}_{\vec{r}+\vec{r}_{\nu},g}+V_{\Uparrow\Uparrow}^{(\nu)}\sum_{\vec{r}}\hat{n}_{\vec{r},\Uparrow}\hat{n}_{\vec{r}+\vec{r}_{\nu},\Uparrow}+V_{\Downarrow\Downarrow}^{(\nu)}\sum_{\vec{r}}\hat{n}_{\vec{r},\Downarrow}\hat{n}_{\vec{r}+\vec{r}_{\nu},\Downarrow}\\
&+V_{g\Uparrow}^{(\nu)}\sum_{\vec{r}}\left(\hat{n}_{\vec{r},g}\hat{n}_{\vec{r}+\vec{r}_{\nu},\Uparrow}+\hat{n}_{\vec{r},\Uparrow}\hat{n}_{\vec{r}+\vec{r}_{\nu},g}\right)+V_{g\Downarrow}^{(\nu)}\sum_{\vec{r}}\left(\hat{n}_{\vec{r},g}\hat{n}_{\vec{r}+\vec{r}_{\nu},\Downarrow}+\hat{n}_{\vec{r},\Downarrow}\hat{n}_{\vec{r}+\vec{r}_{\nu},g}\right)+V_{\Uparrow\Downarrow}^{(\nu)}\sum_{\vec{r}}\left(\hat{n}_{\vec{r},\Uparrow}\hat{n}_{\vec{r}+\vec{r}_{\nu},\Downarrow}+\hat{n}_{\vec{r},\Downarrow}\hat{n}_{\vec{r}+\vec{r}_{\nu},\Uparrow}\right)\\
&+V_{g\Uparrow\Uparrow g}^{(\nu)}\sum_{\vec{r}}\left(\hat{c}_{\vec{r},g}^{\dagger}\hat{c}_{\vec{r},\Uparrow}\hat{c}_{\vec{r}+\vec{r}_{\nu},\Uparrow}^{\dagger}\hat{c}_{\vec{r}+\vec{r}_{\nu},g}+h.c.\right)+V_{g\Downarrow \Downarrow g}^{(\nu)}\sum_{\vec{r}}\left(\hat{c}_{\vec{r},g}^{\dagger}\hat{c}_{\vec{r},\Downarrow}\hat{c}_{\vec{r}+\vec{r}_{\nu},\Downarrow}^{\dagger}\hat{c}_{\vec{r}+\vec{r}_{\nu},g}+h.c.\right)\\
&+V_{\Uparrow\Downarrow\Downarrow\Uparrow}^{(\nu)}\sum_{\vec{r}}\left(\hat{c}_{\vec{r},\Uparrow}^{\dagger}\hat{c}_{\vec{r},\Downarrow}\hat{c}_{\vec{r}+\vec{r}_{\nu},\Downarrow}^{\dagger}\hat{c}_{\vec{r}+\vec{r}_{\nu},\Uparrow}+h.c.\right)+V_{\Uparrow\Downarrow\Uparrow\Downarrow}^{(\nu)}\sum_{\vec{r}}\left(\hat{c}_{\vec{r},\Uparrow}^{\dagger}\hat{c}_{\vec{r},\Downarrow}\hat{c}_{\vec{r}+\vec{r}_{\nu},\Uparrow}^{\dagger}\hat{c}_{\vec{r}+\vec{r}_{\nu},\Downarrow}+h.c.\right) \bigg].
\end{aligned}
\end{equation}
\normalsize
The outer sum over $\nu = X,Y$ corresponds to the nearest-neighbour $X$, $Y$ direction terms respectively. Fig.~\ref{fig_InteractionParameters}(b) shows the nearest-neighbour parameters along the $X$ direction for all of the above terms. They all decay exponentially like the tunneling rates. Unsurprisingly, the terms involving the $\Uparrow$ orbital have the strongest magnitudes since the $\Uparrow$ spatial wavefunction is more delocalized along $X$. The $V_{\Uparrow\Uparrow}^{(X)} = 4 U_{\vec{r},\vec{r}+\vec{r}_X, \vec{r}+\vec{r}_X,\vec{r}}^{\Uparrow\Uparrow\Uparrow\Uparrow}$ interaction of two $\Uparrow$ atoms is the strongest, followed by the interaction $V_{g\Uparrow}^{(X)}=4 U_{\vec{r},\vec{r}+\vec{r}_X, \vec{r}+\vec{r}_X,\vec{r}}^{g\Uparrow\Uparrow g}$ between $\Uparrow$, $g$ atoms and interaction  $V_{\Uparrow\Downarrow}^{(X)}=4 U_{\vec{r},\vec{r}+\vec{r}_X, \vec{r}+\vec{r}_X,\vec{r}}^{\Uparrow\Downarrow\Downarrow\Uparrow}$ between $\Uparrow$, $\Downarrow$ atoms. All of the other terms are smaller by about an order of magnitude, and we neglect them. For the $Y$-direction interactions, we analogously keep only the strongest terms, which will involve the $\Downarrow$ orbital atoms instead. We are left with
\begin{equation}
\begin{aligned}
\hat{H}_{\mathrm{int}}^{\mathrm{(NN)}}&\approx\sum_{\vec{r}}\left[V_{\Uparrow\Uparrow}^{(X)} \hat{n}_{\vec{r},\Uparrow}\hat{n}_{\vec{r}+\vec{r}_X,\Uparrow}+V_{\Uparrow\Downarrow}^{(X)}\left(\hat{n}_{\vec{r},\Uparrow}\hat{n}_{\vec{r}+\vec{r}_X,\Downarrow}+\hat{n}_{\vec{r},\Downarrow}\hat{n}_{\vec{r}+\vec{r}_X,\Uparrow}\right)+V_{g\Uparrow}^{(X)}\left(\hat{n}_{\vec{r},g}\hat{n}_{\vec{r}+\vec{r}_X,\Uparrow} + \hat{n}_{\vec{r},\Uparrow}\hat{n}_{\vec{r}+\vec{r}_X,g}\right)\right]\\
&+\sum_{\vec{r}}\left[V_{\Downarrow\Downarrow}^{(Y)} \hat{n}_{\vec{r},\Downarrow}\hat{n}_{\vec{r}+\vec{r}_Y,\Downarrow}+V_{\Uparrow\Downarrow}^{(Y)}\left(\hat{n}_{\vec{r},\Uparrow}\hat{n}_{\vec{r}+\vec{r}_Y,\Downarrow}+\hat{n}_{\vec{r},\Downarrow}\hat{n}_{\vec{r}+\vec{r}_Y,\Uparrow}\right)+V_{g\Downarrow}^{(Y)}\left(\hat{n}_{\vec{r},g}\hat{n}_{\vec{r}+\vec{r}_Y,\Downarrow}+\hat{n}_{\vec{r},\Downarrow}\hat{n}_{\vec{r}+\vec{r}_Y,g}\right)\right].
\end{aligned}
\end{equation}
As in the on-site case, we can make a further simplification by considering the ground band atoms as frozen, via Eq.~\eqref{eq_GroundBandFrozen}. This causes the $V_{g\Uparrow}^{(X)}$ and $V_{g\Downarrow}^{(Y)}$ terms to also add up to an operator proportional to the total excited atom number, and thus act as an overall constant that can be dropped. Finally, we recognize that for equal lattice depths we have $V_{\Uparrow\Uparrow}^{(X)} = V_{\Downarrow\Downarrow}^{(Y)}$ and $V_{\Uparrow\Downarrow}^{(X)} = V_{\Uparrow\Downarrow}^{(Y)}$. We define,
\begin{equation}
    V_{ee}\equiv V_{\Uparrow\Uparrow}^{(X)}=V_{\Downarrow\Downarrow}^{(Y)},\>\>\>\>\>\>\>\>
    V_{\Uparrow\Downarrow}\equiv V_{\Uparrow\Downarrow}^{(X)} = V_{\Uparrow\Downarrow}^{(Y)}\>\>\>\>\>\>\>\>\text{for }V_X=V_Y,
\end{equation}
which allows $\hat{H}_{\mathrm{int}}^{\mathrm{(NN)}}$ to match the result in the main text the main text.

\section{Bragg dressing}
\label{app_Dressing}
\renewcommand{\theequation}{B\arabic{equation}}
\setcounter{equation}{0}

\subsection{Laser coupling}
Here we detail the implementation of Bragg laser coupling between our desired band states. We write the Hamiltonian of a single lattice site, ignoring interaction effects and assuming a single-particle basis where the excited band states can be treated as first-quantized energy levels $\ket{\Uparrow}$ and $\ket{\Downarrow}$. We also include a manifold of intermediate excited states $\{\ket{E}\}$, which is typically the continuum of an untrapped excited electronic state hyperfine level, separated from the ground electronic states by an optical frequency. We couple the desired band states to the intermediate excited states with two Bragg-type laser beams with single-photon Rabi frequencies $\Omega_1,\Omega_2$ (assumed real for simplicity) and wavevectors $\vec{k}_1,\vec{k}_2$. Assuming that we have a large detuning $\Delta \gg \Omega_1, \Omega_2$ from the intermediate state manifold, the second-order effective coupling between our excited band states $\ket{a},\ket{b} \in \{\ket{\Uparrow},\ket{\Downarrow}\}$ is
\begin{equation}
\begin{aligned}
\bra{a}\hat{H}_{\Omega}\ket{b} \approx -\frac{1}{4\Delta}\sum_{\ket{E}} \bigg[&\Omega_1^2 \bra{a}e^{i \vec{k}_1 \vec{r}} \ket{E}\bra{E} e^{-i \vec{k}_1 \vec{r}}\ket{b} + \Omega_1 \Omega_2 \bra{a}e^{i \vec{k}_1 \vec{r}} \ket{E}\bra{E} e^{-i \vec{k}_2 \vec{r}}\ket{b} \\
+&\Omega_2 \Omega_1 \bra{a}e^{i \vec{k}_2 \vec{r}} \ket{E}\bra{E} e^{-i \vec{k}_1 \vec{r}}\ket{b} +\Omega_2^2 \bra{a}e^{i \vec{k}_2 \vec{r}} \ket{E}\bra{E} e^{-i \vec{k}_2 \vec{r}}\ket{b} \bigg],
\end{aligned}
\end{equation}
where $\vec{r}$ is the position vector from the center of the lattice site. This description is valid provided that the effective bandwidth of the intermediate state manifold $\{\ket{E}\}$ is small compared to $\Delta$. Furthermore, under the assumption that the intermediate state is not trapped by the lattice, it acts as a continuum, allowing the approximation of
\begin{equation}
    \sum_{\ket{E}}\ket{E}\bra{E} = \mathbbm{1}.
\end{equation}
Our matrix elements then simplify to
\begin{equation}
\bra{a}\hat{H}_{\Omega}\ket{b} \approx -\frac{1}{4\Delta}\bigg[\Omega_1^2 \delta_{ab} + \Omega_1 \Omega_2 \bra{a}e^{i (\vec{k}_1-\vec{k}_2)\vec{r}} \ket{b} +\Omega_2 \Omega_1 \bra{a}e^{-i (\vec{k}_1 - \vec{k}_2) \vec{r}}\ket{b} +\Omega_2^2 \delta_{ab}\bigg].
\end{equation}
The first and last term correspond to overall Stark shifts that are equal for both spin states $\ket{\Uparrow}$, $\ket{\Downarrow}$, contributing no overall effect. The middle two terms will create the desired two-photon coupling between the spin states that we are after. The effective two-photon Rabi frequency will be
\begin{equation}
\label{eq_CouplingMatrixElementBragg}
    \frac{\Omega}{2} = -\frac{\Omega_1 \Omega_2}{2\Delta}\bra{\Uparrow}\cos \left( \Delta \vec{k}\cdot\vec{r}\right) \ket{\Downarrow},
\end{equation}
where $\Delta \vec{k} = \vec{k}_1-\vec{k}_2$ is the differential momentum kick of the two laser beams.

\subsection{Elimination of cross-site effects}

Thus far, we have written the Hamiltonian for a single arbitrary lattice site. In principle the position vector $\vec{r}$ multiplying the laser wavevectors must be written using its full real-space value including the lattice site position. At a given lattice site integer index $(i,j)$, we have
\begin{equation}
\vec{r} =  a(i, j, 0) + (X,Y,Z),
\end{equation}
with $a$ the lattice spacing and $(X,Y,Z)$ the position from the center of the site. The cosine in the effective $\hat{H}_{\Omega}$ matrix elements is then written (for a single 2D $X$-$Y$ lattice plane, assuming that $\Delta \vec{k}$ has no $Z$-component)
\begin{equation}
\cos\left(\Delta \vec{k} \cdot\vec{r}\right) = \cos \left[\Delta \vec{k}\cdot\left(X,Y,Z\right) + a\>\Delta \vec{k} \cdot (i,j,0)\right].
\end{equation}
The first term of the cosine argument is the actual spatially-varying phase across the given lattice site, which will be integrated with the on-site lattice Wannier functions of the different cartesian components $X$, $Y$, $Z$. The second term,
\begin{equation}
    \eta \equiv a\>\Delta \vec{k} \cdot (i,j,0), 
\end{equation}
is an additional phase that corresponds to a spin-orbit coupling effect due to the possibly incommensurate wavelengths of the lattice and drives.
Evaluating our desired matrix element between $\ket{\Uparrow}$ and $\ket{\Downarrow}$ we get
\begin{equation}
\begin{aligned}
\label{eq_AppBraggBandCoupling}
\bra{\Uparrow} \hat{H}_{\mathrm{\Omega}} \ket{\Downarrow} = \frac{\Omega_1 \Omega_2}{2\Delta}\cos(\eta)\int_{-\infty}^{\infty} dX\> w_1^{X}(X) \sin\left(\Delta \vec{k}_X X\right) w_0^{X}(X)  \int_{-\infty}^{\infty} dY\> w_0^{Y}(Y) \sin\left(\Delta \vec{k}_Y Y\right) w_1^{Y}(Y),
\end{aligned}
\end{equation}
where $\Delta \vec{k} = (\Delta \vec{k}_{X}, \Delta \vec{k}_{Y}, 0)$ (still assuming no momentum kick along $Z$). The overall two-photon frequency $\Omega$ gets normalized by the spin-orbit coupling phase via the $\cos(\eta)$ prefactor that varies across different lattice sites,
\begin{equation}
\Omega \to \Omega \cos(\eta).
\end{equation}
The Bragg beams effectively act as an additional lattice potential whose direction and wavelength depends on the differential momentum kick $\Delta \vec{k}$, only providing a coupling between motional states rather than a Stark shift.

Ideally we want to have as large and as uniform of a Rabi frequency $\Omega$ as possible across the lattice. Making it large requires both $\Delta \vec{k}_X$ and $\Delta \vec{k}_Y$ to be non-zero, as otherwise the sines in the integrands above will vanish due to the Wannier functions' even/odd spatial symmetry. For example, we could use the scheme in the main text, where one beam comes in along the $X$ direction and one along the $Y$ direction, each co-propagating with the lattice beams,
\begin{equation}
\label{eq_ExampleWavevectorBragg}
    \text{Example: }\>\>\>\>\>\vec{k}_1 = \frac{2\pi}{\lambda}(1,0,0),\>\>\>\vec{k}_2 = \frac{2\pi}{\lambda} (0,-1,0),\>\>\> \to \>\>\> \Delta \vec{k} = \frac{2\pi}{\lambda} (1,1,0),
\end{equation}
with $\lambda$ the wavelength of the Bragg beams. On the other hand, making the Rabi frequency uniform requires us to minimize the effect of the spin-orbit coupling phase. This can be done by making the phase as close as possible to an integer multiple of $2\pi$,
\begin{equation}
    \eta =2\pi m,\>\> m\in \mathbbm{Z},
\end{equation}
which requires that both the $X$ and $Y$ directions independently satisfy
\begin{equation}
a \Delta \vec{k}_X = 2\pi m, \>\>\> a \Delta \vec{k}_{Y} = 2\pi m',\>\>\>m,m' \in \mathbbm{Z} \,.
\end{equation}
For our example, this condition amounts to
\begin{equation}
    \text{Example: }\>\>\>\>\>\frac{2\pi a}{\lambda} = 2\pi m,\>\>\> m \in \mathbbm{Z},
\end{equation}
which can be satisfied with an appropriate choice of wavelengths and/or lattice spacing such as by choosing $\lambda = a$,  i.e Bragg beams with half the lattice beam wavelength.

\subsection{SOC Benchmarking}

\begin{figure}[tb]
\centering
\includegraphics[width=0.5\linewidth]{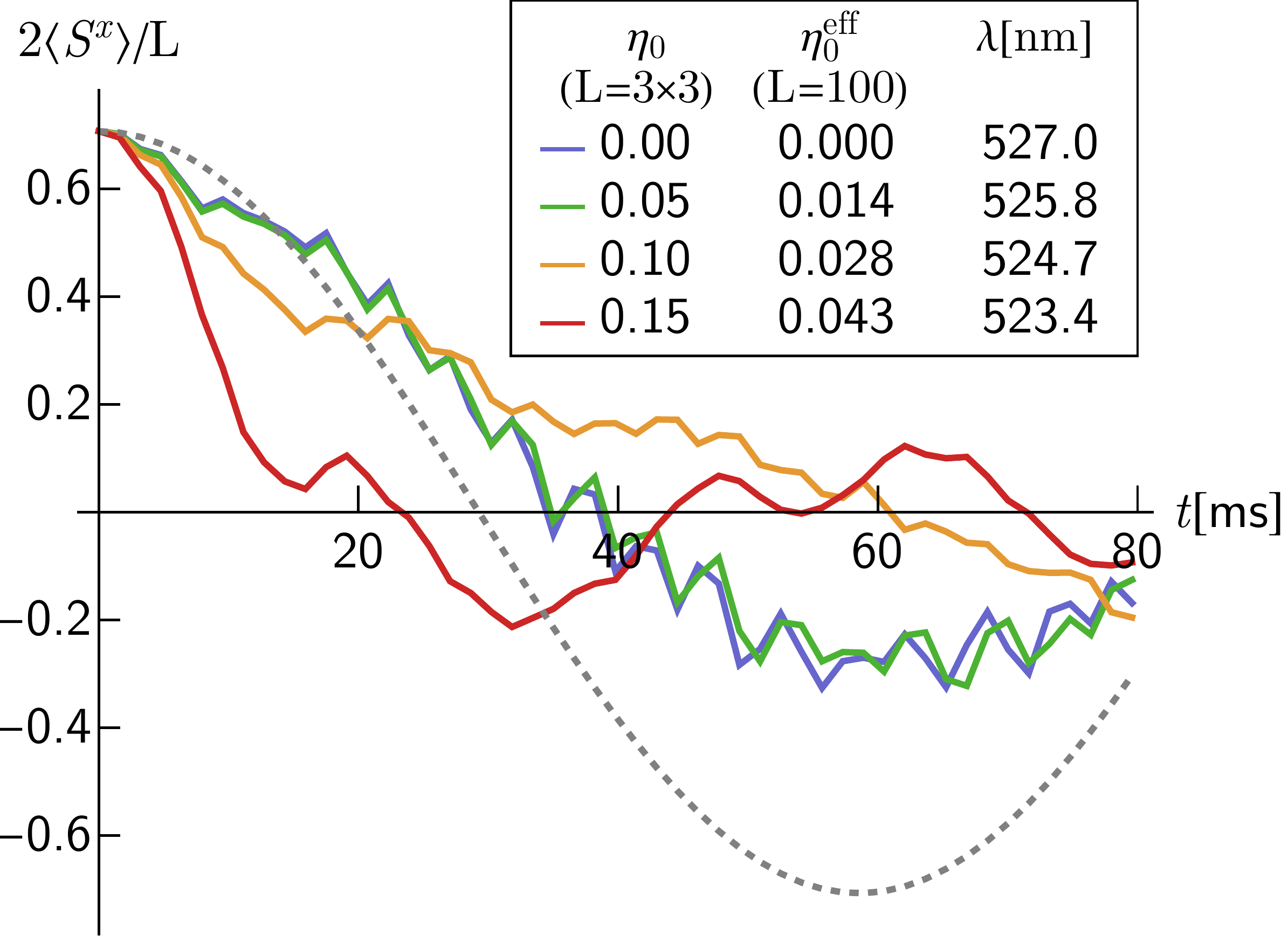}
\caption{Time-evolution of $\langle\hat{S}^{x}\rangle$ to measure the mean-field density phase shift, under the presence of spin-orbit coupling effects. We simulate an $L=3\times 3$ system using the Fermi-Hubbard model, written in real space with periodic boundary conditions, starting from a collective product state $\ket{\psi_0}$ inclined at $\theta = \pi/4$ as per the protocol described in the main text. The gray dashed line is the expected mean-field precession. The drive on each lattice site $(i,j)$ is modulated by the presence of spin-orbit coupling, $\Omega \to \Omega \cos(\eta)$ with $\eta = \eta_0 [(i-i_0) + (j-j_0)]$ [with $(i_0, j_0)$ the center site of the lattice, here $(2,2)$]. Different color dots represent the evolution for different values of $\eta_0$. The legend also shows the effective $\eta_0^{\mathrm{eff}}$ for a larger system of size $L= 100$, calculated by computing the average value of $\cos(\eta)$ for a given $\eta_0$, and determining what the corresponding $\eta_0^{\mathrm{eff}}$ would need to be to have the same average if sampling over a $10 \times 10$ grid with it. The corresponding laser drive wavelength $\lambda$ is also provided, assuming lattice period $a = 527$\,nm, and Bragg beams co-propagating with the $X$, $Y$ lattice beams respectively.}
\label{fig_SOC}
\end{figure}

We benchmark how much of a dressing laser wavelength mismatch can be tolerated while still maintaing a coherent collective-spin signal in the Ramsey spectroscopy. We assume the drive lasers to co-propagate with the $X$, $Y$ lattice axes as in the example above. The wavelength mismatch is quantified by a dimensionless parameter $\eta_0$
\begin{equation}
\eta_0 \equiv \frac{2\pi a}{\lambda} - 2\pi \,.
\end{equation}
When $\eta_0 = 0$, the SOC phase is always an integer multiple of $2\pi$ and there are no SOC modulation effects on the drive Rabi frequency $\Omega$. As $\eta_0$ increases, the modulation gets stronger, and the coherence of the collective spin signal is lost. In Fig.~\ref{fig_SOC} we show the time-evolution of the density phase via $\langle \hat{S}^{x} \rangle$ for different values of $\eta_0$. We assume that the maximum-amplitude region is at the center of the lattice, for which we would have
\begin{equation}
    \eta = \eta_0 [(i-i_0) + (j-j_0)],
\end{equation}
with $(i_0,j_0)$ the central site of the lattice. The simulation uses a small system size $L = 3\times 3$ with periodic boundary conditions (while these are somewhat unphysical given the spatial structure $\Omega\cos(\eta)$ of the modulation, they help avoid strong boundary effects). We see that for sufficiently small $\eta_0$ mismatch the signal's oscillation correctly follows the mean-field envelope (up to decay of the contrast). A larger $\eta_0$ causes the signal oscillation period to change, because there are additional perturbative contributions to the mean-field precession rate $\chi$ that emerge. Even larger $\eta_0$ will cause the signal to decay altogether as the gap protection keeping the system in the highest angular momentum shell breaks down.

While the modulation grows stronger for every lattice site increment away from the center, we can still estimate the maximum tolerable mismatch for a larger lattice size. A larger $\eta_0$ for our $L=3 \times 3$ system simply corresponds to a smaller effective $\eta_0^{\mathrm{eff}}$ for a bigger system. In the legend of Fig.~\ref{fig_SOC}, we show what that effective $\eta_0^{\mathrm{eff}}$ would be for an $L = 10 \times 10$ system, by simply computing what average value of $\cos[\eta_0^{\mathrm{eff}} (i-i_0+j-j_0)]$ for $L = 10\times 10$ matches the average value of $\cos[\eta_0 (i-i_0+j-j_0)]$ for $L = 3\times 3$. This $\eta_0^{\mathrm{eff}}$ provides an estimate of what kind of mismatch a realistic experiment can tolerate assuming it traps atoms in a $10\times 10$ site region at the lattice center. The figure also provides the corresponding laser wavelength $\lambda$ for lattice spacing $a = 527$ nm. To have $\eta_0^{\mathrm{eff}}=0$ we would need $\lambda = a$ exactly, and we can tolerate deviations of about $1-2$ nm away from that.

\subsection{Other bands}
Thus far, we have only included the two excited motional bands $\ket{\Uparrow}$ [excitation numbers $(1,0,0)$] and $\ket{\Downarrow}$ [$(0,1,0)$] in our calculations. A 3D optical lattice will have other bands, including both the ground band $g$ [$(0,0,0)$] and higher bands such as $(2,0,0)$, $(0,0,1)$, etc.. While the ground band is filled, the higher bands are not, and can be coupled to by the drive. This sets an upper limit on how high we can make the two-photon Rabi frequency $\Omega$, as increasing it will also increase the couplings to the other bands, which will cause the system to heat once they become comparable to the band-gaps.

To ensure that this does not happen in our system, we simulate the dynamics of a single lattice site now using a two-atom Fock basis, starting with our standard initial condition of one atom in $g$ and one in $\Uparrow$, then including all possible other band states with up to three motional excitations. We then compute all coupling matrix elements induced by our lasers via Eq.~\eqref{eq_CouplingMatrixElementBragg}, where $\ket{a}$, $\ket{b}$ now run over all of the included band states. The wavevectors of the lasers are chosen to co-propagate with the lattice beams as in the example of Eq.~\eqref{eq_ExampleWavevectorBragg}, with a wavelength $\lambda = a$ to avoid any SOC effects. We include only the single-particle terms of the drive and band-gaps without the interactions, as we work in a regime where the drive is the dominant energy scale in the system.

\begin{figure}[tb]
\centering
\includegraphics[width=0.4\linewidth]{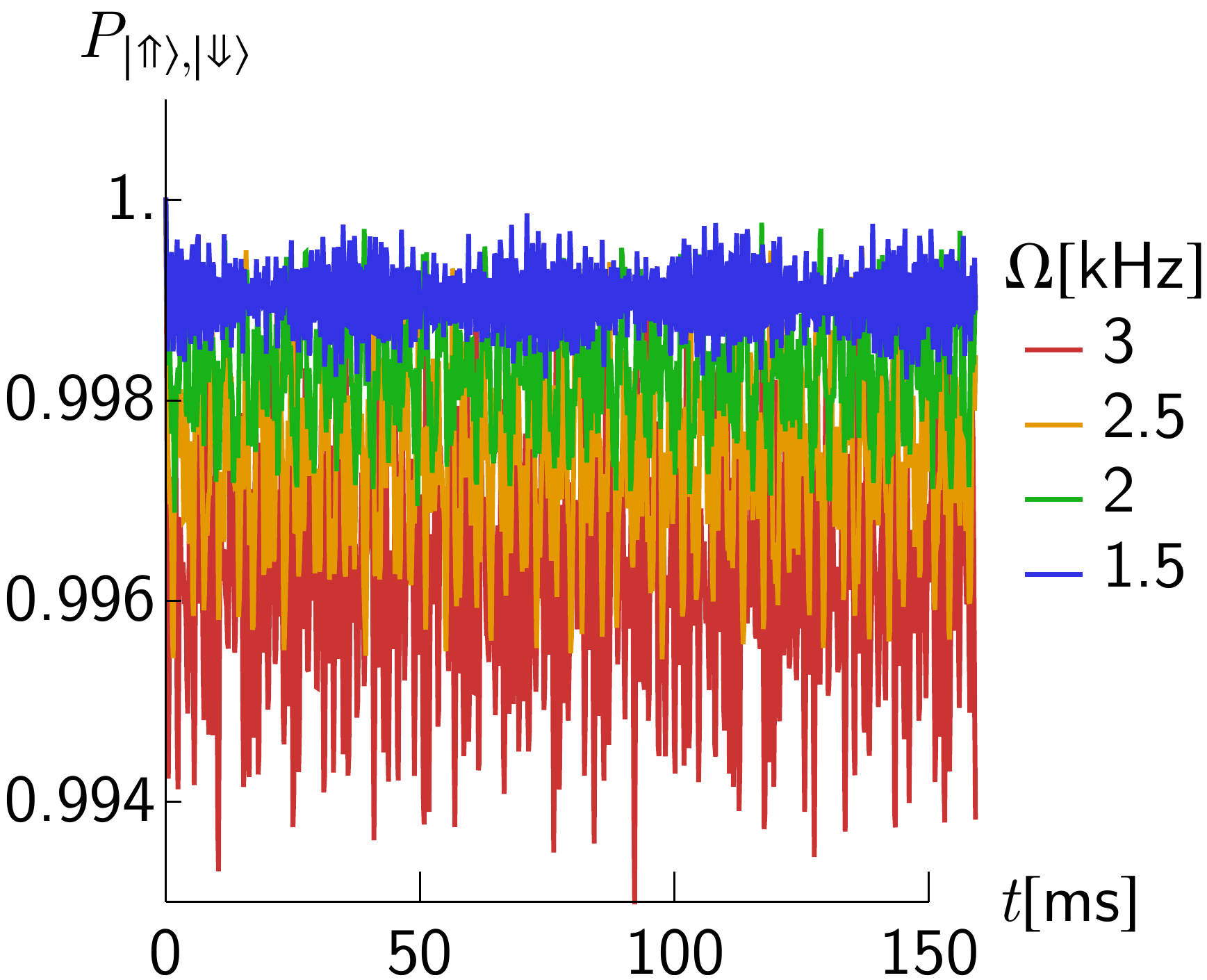}
\caption{Wavefunction population in the desired spin-1/2 subspace for a single lattice site evolving under the laser drive and including higher motional bands. All band states with up to 3 motional excitations are included. Only single-particle terms (the laser drive couplings and the band gaps) are used, as these are the dominant energy scales in the system.}
\label{fig_OtherBands}
\end{figure}

Fig.~\ref{fig_OtherBands} shows the dynamics of wavefunction population $P_{\ket{\Uparrow},\ket{\Downarrow}}$ in the manifold of our two desired spin states, thus the two Fock states both with one atom in $g$, and one atom in $\Downarrow$ or $\ket{\Downarrow}$ respectively, for different values of $\Omega$.

We note that a coherent Bragg laser field is advantageous for suppressing heating because it mitigates couplings to the closest-lying unwanted band states. The coupling between generic orbitals $\ket{a}$ and $\ket{b}$ is an extension of Eq.~\eqref{eq_AppBraggBandCoupling}, which can be written as (still assuming $\Delta k_{Z} =0$)
\begin{equation}
\begin{aligned}
\bra{a} \hat{H}_{\mathrm{\Omega}} \ket{b} = \frac{\Omega_1 \Omega_2}{2\Delta}\cos(\eta)\int_{-\infty}^{\infty} dX\> w_{n_a^{X}}^{X}(X) \sin\left(\Delta \vec{k}_X X\right) w_{n_b^{X}}^{X}(X)  \int_{-\infty}^{\infty} dY\> w_{n_a^{Y}}^{Y}(Y) \sin\left(\Delta \vec{k}_Y Y\right) w_{n_b^{Y}}^{Y}(Y),
\end{aligned}
\end{equation}
with $(n_a^X,n_a^Y,n_a^Z)$ and $(n_b^X, n_b^Y, n_b^Z)$ the orbital harmonic excitation numbers of states $\ket{a}$ and $\ket{b}$. Our desired subspace consists of the states with excitations $(1,0,0)$ and $(0,1,0)$. Assuming a deeper lattice along $Z$, the other closest-lying band states are $(0,0,0)$ (the ground band), $(2,0,0)$, $(0,2,0)$ and $(1,1,0)$ (doubly-excited $X$, $Y$ bands). For all of these, the above coupling to $(1,0,0)$ and $(0,1,0)$ will vanish due to spatial parity. The only way population can leak is through couplings to triply-excited states or above, such as $(3,0,0)$. The energy gap to these is twice the band-gap, which permits a larger value of $\Omega$ without significant population loss.

\section{Spin model derivation using no-mode-changing-collisions approximation}
\label{app_SpinDerivation}
\renewcommand{\theequation}{C\arabic{equation}}
\setcounter{equation}{0}
Here we show how the Fermi-Hubbard model together with the laser drive $\hat{H}_{\mathrm{FH}} + \hat{H}_{\Omega}$ can be approximated with a spin model under a no-mode-changing-collisions approximation. As a demonstrative example, we show how the nearest-neighbour density-density interaction of two $\Uparrow$ atoms is transformed into a spin term. We work in the non-dressed basis $\{\hat{c}_{\vec{r},\Uparrow},\hat{c}_{\vec{r},\Downarrow}\}$, and convert to the dressed basis at the end. The corresponding Fermi-Hubbard interaction term can be written as 
\begin{equation}
    I_{\Uparrow\Uparrow} = V_{ee}\sum_{\vec{r}}\hat{n}_{\vec{r},\Uparrow}\hat{n}_{\vec{r}+\vec{r}_{X},\Uparrow} = \frac{V_{ee}}{L}\sum_{k_X,k_X',q_X}\sum_{k_Y,k_Y',q_Y}e^{-i(k_X - k_X' + q_X)}\hat{c}_{k_X+q_X,k_Y+q_Y,\Uparrow}^{\dagger}\hat{c}_{k_X',k_Y',\Uparrow}\hat{c}_{k_X'-q_X,k_Y'-q_Y,\Uparrow}^{\dagger}\hat{c}_{k_X,k_Y,\Uparrow},
\end{equation}
where $k_X, k_X', q_X$ and $k_Y, k_Y', q_Y$ are 2D quasimomenta along $X$ and $Y$ lattice directions respectively. The no-mode-changing-collisions approximation amounts to only keeping those terms of the sums that conserve individual quasimomentum mode population,
\begin{equation}
\hat{c}_{\vec{k}_1,\Uparrow}^{\dagger}\hat{c}_{\vec{k}_{2},\Uparrow}\hat{c}_{\vec{k}_{3},\Uparrow}^{\dagger}\hat{c}_{\vec{k}_4,\Uparrow} = 0 \>\>\>\text{ unless $\vec{k}_1 = \vec{k}_2$, $\vec{k}_3 = \vec{k}_4$ \textbf{or} $\vec{k}_1 = \vec{k}_4$, $\vec{k}_2 = \vec{k}_3$}.
\end{equation}
Applying this approximation leaves:
\begin{equation}
I_{\Uparrow\Uparrow} = \frac{V_{ee}}{L}\sum_{\vec{k},\vec{k'}}e^{-i(k_X-k_X')}\hat{c}_{\vec{k},\Uparrow}^{\dagger}\hat{c}_{\vec{k}',\Uparrow}\hat{c}_{\vec{k}',\Uparrow}^{\dagger}\hat{c}_{\vec{k},\Uparrow}+\frac{V_{ee}}{L}\sum_{\vec{k}\neq \vec{k}'}\hat{c}_{\vec{k},\Uparrow}^{\dagger}\hat{c}_{\vec{k},\Uparrow}\hat{c}_{\vec{k}',\Uparrow}^{\dagger}\hat{c}_{\vec{k}',\Uparrow},
\end{equation}
using indices $\vec{k} = (k_X,k_Y)$, $\vec{k}'=(k_X',k_Y')$ in this equation and hereafter. We now commute the operators and map them to spin-1/2 operators via
\begin{equation}
\label{eq_NMCAppSpin}
    \hat{c}_{\vec{k},\Uparrow}^{\dagger}\hat{c}_{\vec{k},\Downarrow} =\hat{\sigma}_{\vec{k}}^{'+},\>\>\>\>\hat{c}_{\vec{k},\Uparrow}^{\dagger}\hat{c}_{\vec{k},\Uparrow} = \frac{1}{2}\left(\mathbbm{1} + \hat{\sigma}_{\vec{k}}^{'z}\right),\>\>\>\>\hat{c}_{\vec{k},\Downarrow}^{\dagger}\hat{c}_{\vec{k},\Downarrow} = \frac{1}{2}\left(\mathbbm{1} - \hat{\sigma}_{\vec{k}}^{'z}\right).
\end{equation}
These spin operators are denoted with a prime because we are still working in the un-dressed basis. Applying this mapping to our example term, we get
\begin{equation}
    I_{\Uparrow\Uparrow} = \frac{V_{ee}}{4L}\sum_{\vec{k},\vec{k}'}\left[1-\cos(k_X a-k_X' a)\right]\hat{\sigma}_{\vec{k}}^{'z}\hat{\sigma}_{\vec{k}'}^{'z} + \frac{V_{ee}}{2}\sum_{\vec{k}}\hat{\sigma}_{\vec{k}}^{' z}.
\end{equation}
Finally, we rewrite the model in the dressed basis. Since we have already mapped to a spin model, we simply apply the corresponding rotation to the spin operators,
\begin{equation}
\label{eq_NMCAppRotation}
    \hat{\sigma}_{\vec{k}}^{'x} = \hat{\sigma}_{\vec{k}}^{z},\>\>\>\>\hat{\sigma}_{\vec{k}}^{'y} = -\hat{\sigma}_{\vec{k}}^{y},\>\>\>\>\hat{\sigma}_{\vec{k}}^{'z} = \hat{\sigma}_{\vec{k}}^{x},
\end{equation}
where the non-primed spin operators are the ones used in the main text. We end up with
\begin{equation}
    I_{\Uparrow\Uparrow} = \frac{V_{ee}}{4L}\sum_{\vec{k},\vec{k}'}\left[1-\cos(k_Xa-k_Xa')\right]\hat{\sigma}_{\vec{k}}^{x}\hat{\sigma}_{\vec{k}'}^{x}+ \frac{V_{ee}}{2}\sum_{\vec{k}}\hat{\sigma}_{\vec{k}}^{x}.
\end{equation}
The same procedure is applied to all of the other Fermi-Hubbard interaction terms. Adding them all together yields the spin-spin interactions in the main text Hamiltonian $\hat{H}_{\mathrm{S}}$.

The single-particle tunneling and drive terms can also be directly written as effective spin rotations by writing them in momentum space, mapping to spin operators via Eq.~\eqref{eq_NMCAppSpin}, then rotating into the dressed spin basis via Eq.~\eqref{eq_NMCAppRotation}. We have the tunneling of the excited atoms, which is written as (starting from position space)
\begin{equation}
\begin{aligned}
    \hat{H}_{J} &= \sum_{\vec{r}} \left(-J_0 \hat{c}_{\vec{r},\Downarrow}^{\dagger}\hat{c}_{\vec{r}+\vec{r}_X,\Downarrow} + J_1 \hat{c}_{\vec{r},\Uparrow}^{\dagger}\hat{c}_{\vec{r}+\vec{r}_X,\Uparrow} - J_0 \hat{c}_{\vec{r},\Uparrow}^{\dagger}\hat{c}_{\vec{r}+\vec{r}_Y,\Uparrow} + J_1 \hat{c}_{\vec{r},\Downarrow}^{\dagger}\hat{c}_{\vec{r}+\vec{r}_Y,\Downarrow} + h.c.\right)\\
    &= \sum_{\vec{k}} \left[-2J_0 \cos(k_Xa)\hat{n}_{\vec{k},\Downarrow} + 2J_1\cos(k_Xa) \hat{n}_{\vec{k},\Uparrow}- 2J_0\cos(k_Ya) \hat{n}_{\vec{k},\Uparrow} + 2J_1 \cos(k_Ya)\hat{n}_{\vec{k},\Downarrow}\right]\\
    &=(J_1+J_0)\sum_{\vec{k}} [\cos(k_Xa)-\cos(k_Ya)](\hat{n}_{\vec{k},\Uparrow} -\hat{n}_{\vec{k},\Downarrow})+(J_1-J_0) \sum_{\vec{k}}[\cos(k_Xa)+\cos(k_Ya)](\hat{n}_{\vec{k},\Uparrow}+\hat{n}_{\vec{k},\Downarrow})\\
    &=\sum_{\vec{k}} \epsilon_{\vec{k}}(\hat{n}_{\vec{k},\Uparrow} -\hat{n}_{\vec{k},\Downarrow})+ \sum_{\vec{k}}\overline{E}_{\vec{k}}(\hat{n}_{\vec{k},\Uparrow}+\hat{n}_{\vec{k},\Downarrow})\\
    &= \sum_{\vec{k}}\epsilon_{\vec{k}}\hat{\sigma}_{\vec{k}}^{x} + \sum_{\vec{k}}\overline{E}_{\vec{k}}\mathbbm{1},
\end{aligned}
\end{equation}
with the parameters $\epsilon_{\vec{k}} = (J_1 + J_0)[\cos(k_Xa) - \cos(k_Ya)]$ and $\overline{E}_{\vec{k}} = (J_1 - J_0)[\cos(k_Xa) + \cos(k_Ya)]$ defined as in the main text. We also have the drive,
\begin{equation}
\begin{aligned}
    \hat{H}_{\Omega} &= \frac{\Omega}{2}\sum_{\vec{k}} \left(\hat{c}_{\vec{k},\Uparrow}^{\dagger}\hat{c}_{\vec{k},\Downarrow}+h.c.\right)= \frac{\Omega}{2}\sum_{\vec{k}}\hat{\sigma}_{\vec{k}}^{z}.
\end{aligned}
\end{equation}

\section{State preparation}
\label{app_StatePreparation}
\renewcommand{\theequation}{D\arabic{equation}}
\setcounter{equation}{0}

\begin{figure}[tb]
\centering
\includegraphics[width=0.3\linewidth]{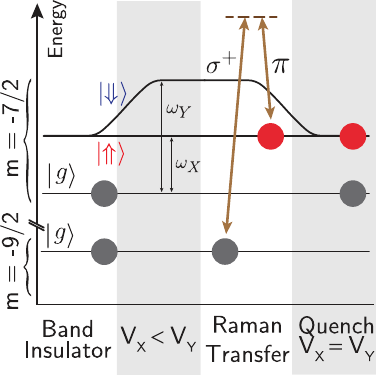}
\caption{Protocol for preparing the initial state $\ket{\psi_0}$ with no tipping angle $\theta=0$. Using ultracold $^{40}$K atoms as an example, one starts with a band insulator of two ground band atoms per site in two nuclear-spin states $\ket{m = -9/2}$, $\ket{m = -7/2}$ of the ground hyperfine manifold. The lattice depths are set to $V_X < V_Y$, causing the $\ket{\Uparrow}$ state to be lower in energy than $\ket{\Downarrow}$. The $\ket{-9/2,g}$ ground-band atom is transferred to the $\ket{-7/2,\Uparrow}$ state using a Raman transition with one linearly- and one circularly-polarized beam through an intermediate state such as the electronic $^{2}P_{1/2}$, $\ket{F=7/2,m = -7/2}$ state. The $V_Y$ depth is then quenched back to match $V_X$.}
\label{fig_StatePreparation}
\end{figure}

Here we discuss experimentally-realistic methods for preparing the initial state $\ket{\psi}_0=e^{i \theta \hat{S}^{y}}\prod_{\vec{k}}\ket{\rightarrow}_{\vec{k}}$. The first step is to prepare a state with all spins pointing along the $x$ direction of the dressed Bloch sphere ($\theta = 0$). A protocol for of this state is depicted in Fig.~\ref{fig_StatePreparation}. One starts from a standard band insulator with double occupancy in the ground band $g$, using the two lowest-energy hyperfine states of the ground electronic state. The lattice depths along $X$ and $Y$ are made unequal with $V_{Y} > V_{X}$, so that the $\ket{\Downarrow}$ state is higher in energy than $\ket{\Uparrow}$ ($V_Z$ remains much stronger than $V_{X}$ as well). A spin-polarized two-band insulator is created with a sideband-resolved Raman pulse, transferring the $g$ band atom of one hyperfine state into the $\ket{\Uparrow}$ band of the other hyperfine state. The $V_Y$ lattice depth is then quenched to match $V_X$, which brings the system to the configuration discussed in the main text.

Generating a tipping angle $\theta$ may be done by enabling the Bragg dressing beams $\hat{H}_{\Omega}$ after the protocol above, and allowing them to run for a time $t \Omega = \theta$. This executes a $\theta$-pulse rotating the collective spin away from the equator. One must then advance the phase of the two-photon Bragg coupling ahead by $\pi/2$, with e.g. a fast pulse of the Bragg laser detunings. In the dressed frame where the drive is pointed along the $z$-axis of the Bloch sphere after this phase advancement, the resulting state will now take the desired form $\ket{\psi_0}$ with the tipping angle $\theta$ in the $x$-$z$ plane.

\section{Imperfect filling fraction}
\label{app_Filling}
\renewcommand{\theequation}{E\arabic{equation}}
\setcounter{equation}{0}

\begin{figure}[tb]
\centering
\includegraphics[width=0.7\linewidth]{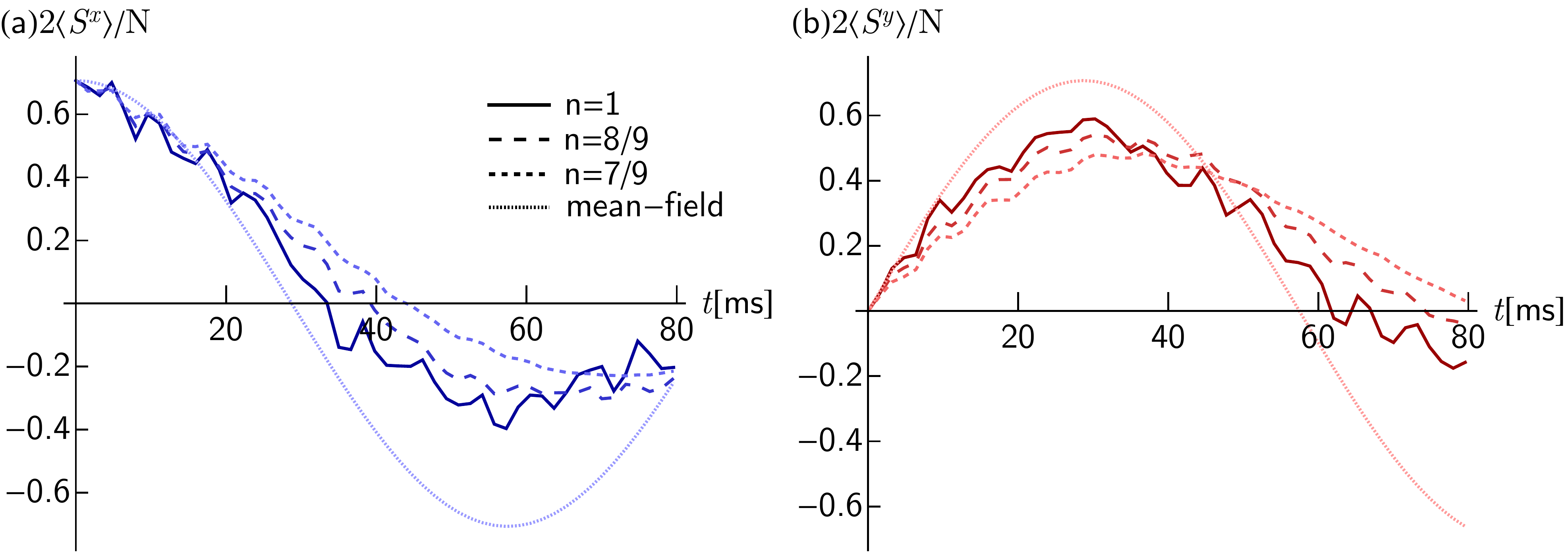}
\caption{Density phase shift precession measured by (a) $\langle\hat{S}^{x} \rangle$ and (b) $\langle\hat{S}^{y} \rangle$ for $L = 3 \times 3$ systems with holes in the initial loadout. We assume all lattice sites have a $g$ atom, but that some are missing an excited-band atom acting as the spin-1/2. The $n=1$ has ideal filling, while $n=8/9$ has no excited atom at lattice site $(i,j)=(3,3)$ (using periodic boundaries), and $n=7/9$ no excited atoms at $(3,3)$ and $(1,2)$. The mean-field lines show the ideal oscillation envelopes with no contrast decay (the expected behaviour in the $L = N \to \infty$ limit).}
\label{fig_Filling}
\end{figure}

The energetic gap of a band insulator has allowed for fidelity of over 99\% in the central ten-by-ten core of a two-dimensional optical lattice~\cite{Chiu:2018fl}. The remaining holes will introduce errors in our protocol, as they facilitate mode-changing collisions that invalidate the spin model approximation. However, with a sufficiently low defect density, we still expect to observe a collective-spin signal from the p-wave interactions. Figure~\ref{fig_Filling} shows the density phase precession of the collective spin using the protocol in the main text, for initial conditions that have one or more holes in place of the excited atoms (still assuming a filled ground band). For a sufficiently low hole fraction, the amplitude of oscillation is somewhat reduced, but one can still infer the oscillation period set by $\chi$.

\section{Collective spin model derivation}
\label{app_CollectiveDerivation}
\renewcommand{\theequation}{F\arabic{equation}}
\setcounter{equation}{0}

\subsection{Collective basis}
Here we show how the spin model $\hat{H}_{\mathrm{S}}$ in the main text can be transformed into a collective one-axis twisting model. We start by splitting the spin model into pieces:
\begin{equation}
\begin{aligned}
\hat{H}_{\mathrm{S}} &= \hat{H}_0 + \hat{H}_{\mathrm{int}}^{\mathrm{(NN)}}+ \hat{H}_{J},\\
\hat{H}_{0}&=\hat{H}_{\mathrm{int}}^{\mathrm{(OS)}} + \hat{H}_{\Omega}.
\end{aligned}
\end{equation}
The first piece $\hat{H}_0$ consists of the Heisenberg term $\sim U_{\Uparrow\Downarrow}$ from the on-site interactions and the drive $\sim \Omega$,
\begin{equation}
\hat{H}_0 = -\frac{U_{\Uparrow\Downarrow}}{4L}\sum_{\vec{k},\vec{k}'}\vec{\sigma}_{\vec{k}}\cdot\vec{\sigma}_{\vec{k}'} + \frac{\Omega}{2}\sum_{\vec{k}}\hat{\sigma}_{\vec{k}}^{z}.
\end{equation}
The second piece consists of the nearest-neighbour interactions, which can be written as
\begin{equation}
\begin{aligned}
\hat{H}_{\mathrm{int}}^{\mathrm{(NN)}} &=\frac{1}{4L}\sum_{\vec{k},\vec{k}'}\left[2V_{ee}-4V_{\Uparrow\Downarrow}-V_{ee}\left(\cos(k_Xa-k_X'a)+\cos(k_Ya-k_Y'a)\right)\right]\hat{\sigma}_{\vec{k}}^{x}\hat{\sigma}_{\vec{k}'}^{x}\\
&-\frac{V_{\Uparrow\Downarrow}}{2L}\sum_{\vec{k},\vec{k}'}\left[\cos(k_Xa-k_X'a)+\cos(k_Ya-k_Y'a)\right]\left(\hat{\sigma}_{\vec{k}}^{y}\hat{\sigma}_{\vec{k}'}^{y}+\hat{\sigma}_{\vec{k}}^{z}\hat{\sigma}_{\vec{k}'}^{z}\right),
\end{aligned}
\end{equation}
and the third piece is the single-particle terms coming from the tunneling,
\begin{equation}
\begin{aligned}
\hat{H}_{J} &=(J_1+J_0)\sum_{\vec{k}}\left[\cos(k_Xa)-\cos(k_Ya)\right]\hat{\sigma}_{\vec{k}}^{x}.
\end{aligned}
\end{equation}
The first piece $\hat{H}_0$ contains the energetically-strongest terms, which we can write as a collective-spin Hamiltonian,
\begin{equation}
\hat{H}_0 = -\frac{U_{\Uparrow\Downarrow}}{L} \vec{S} \cdot \vec{S} + \Omega \hat{S}^{z},
\end{equation}
with $\vec{S} = (\hat{S}^{x},\hat{S}^{y},\hat{S}^{z})$. As discussed in the main text, this Hamiltonian conserves total angular momentum $S$ and creates an energy gap between the fully-symmetric Dicke manifold $S=L/2$ and the next shell $S = L/2-1$. We can thus approximate the effect of the other two pieces $\hat{H}_{\mathrm{int}}^{\mathrm{(NN)}}$, $\hat{H}_{J}$ on unitary evolution of fully-symmetric product states by projecting these pieces into the Dicke manifold via perturbation theory.

\subsection{Nearest-neighbour interaction terms}
The nearest-neighbour interactions $\hat{H}_{\mathrm{int}}^{\mathrm{(NN)}}$ are easiest to start with because their dominant contribution is just their projection into the Dicke manifold directly. These interactions consist of two-body spin terms $\hat{\sigma}_{\vec{k}}^{\alpha} \hat{\sigma}_{\vec{k}'}^{\alpha}$ (for $\alpha \in \{x,y,z\}$), which can be turned into collective terms via
\begin{equation}
\hat{\sigma}_{\vec{k}}^{\alpha} \hat{\sigma}_{\vec{k}'}^{\alpha} \to (1-\delta_{\vec{k},\vec{k}'}) \frac{4}{L(L-1)}\hat{S}^{\alpha} \hat{S}^{\alpha},
\end{equation}
which amounts to ignoring the spin quasi-momentum indices unless the term involves a product of two equal spin operators in the same quasi-momentum mode, in which case the term equals the identity and is ignored. We divide by $L(L-1)$ since the all-to-all $\hat{S}^{\alpha}\hat{S}^{\alpha}$ consists of $L(L-1)$ non-identity interaction terms compared to a single one, and multiply by 4 because of the factor of two difference between Pauli and spin operators. The momentum-dependent coefficients of the individual terms are averaged, causing any cosine terms to vanish, leaving
\begin{equation}
\hat{H}_{\mathrm{int}}^{\mathrm{(NN)}} \to  \frac{2}{L}\left(V_{ee} - 2 V_{\Uparrow\Downarrow}\right) \hat{S}^{x}\hat{S}^{x}.
\end{equation}
This collective term represents the dominant effect of the nearest neighbour interactions in the Dicke manifold. While there can also be higher order effects from the nearest-neighbour terms, we ignore them because the overall nearest-neighbour interaction coefficients are already small compared to the energy gaps set by $\hat{H}_0$.

\subsection{Tunneling terms}
We now turn to the tunneling terms $\hat{H}_{J}$. We can likewise project these into the Dicke manifold by replacing $\hat{\sigma}_{\vec{k}}^{x}$ with $\frac{2}{L}\hat{S}^{x}$. Because of the cosine prefactors, however, this projection will average to zero and the tunneling terms will have no zeroth-order effect. Their lowest-order contribution will instead come in as a second-order perturbation.

The standard second-order Schrieffer-Wolff correction to the Hamiltonian can be written as
\begin{equation}
\hat{H}_{J,\mathrm{eff}} = -\frac{1}{2}[\mathcal{O}\hat{H}_{J}, \mathcal{L}\hat{H}_{J}],
\end{equation}
where $\mathcal{O}\hat{H}_{J}$ consists of all matrix elements of $\hat{H}_{J}$ coupling \textit{different} angular momentum $S$-shells, and
\begin{equation}
\mathcal{L}\hat{H}_{J} = \sum_{\alpha,\beta} \frac{1}{E_{\alpha} - E_{\beta}} \ket{\alpha}\bra{\alpha}\mathcal{O}\hat{H}_{J} \ket{\beta}\bra{\beta},
\end{equation}
where the sums $\alpha$, $\beta$ run over all states in the Hilbert space and $E_{\alpha} = \bra{\alpha} \hat{H}_{0} \ket{\alpha}$, $E_{\beta}=\bra{\beta}\hat{H}_{0}\ket{\beta}$ are unperturbed state energies.

Since we are interested in dynamics that start with collective product states, to good approximation we can restrict our analysis to the Dicke manifold where the product states reside ($S = L/2$) and the next-lowest angular momentum shell of spin-waves ($S=L/2 - 1$). The Dicke states will be written in shorthand as $\ket{S=L/2,m} = \ket{m}$ for $m \in S,S-1, \dots, -S$. The spin-wave states are labelled by an additional index $k \in 1 \cdots L-1$, written in shorthand as $\ket{S=L/2-1,m,k} = \ket{mk}$. We write our perturbation theory terms using this shorthand:
\begin{equation}
\mathcal{O}\hat{H}_{J} = \sum_{m=-L/2}^{L/2} \sum_{m' = -L/2+1}^{L/2-1} \sum_{k=1}^{L-1} \left(\ket{m}\bra{m'k} \bra{m} \hat{H}_{J}\ket{m'k} + h.c.\right),
\end{equation}
and
\begin{equation}
\mathcal{L}\hat{H}_{J} = \sum_{m=-L/2}^{L/2} \sum_{m' = -L/2+1}^{L/2-1} \sum_{k=1}^{L-1}\frac{1}{-U_{\Uparrow\Downarrow} + \Omega(m-m')} \left(\ket{m}\bra{m'k} \bra{m} \hat{H}_{J}\ket{m'k} - h.c.\right).
\end{equation}
We evaluate the commutator of these two, yielding
\footnotesize
\begin{equation}
\begin{aligned}
\hat{H}_{J,\mathrm{eff}} &= \frac{1}{2} \sum_{m=-L/2}^{L/2}\sum_{\tilde{m}=-L/2}^{L/2}\sum_{m' = -L/2+1}^{L/2-1} \sum_{k=1}^{L}\frac{1}{-U_{\Uparrow\Downarrow}+ \Omega(m-m')} \left(\ket{\tilde{m}}\bra{m} \> \bra{\tilde{m}}\hat{H}_{J}\ket{m'k}\bra{m'k}\hat{H}_{J} \ket{m} +h.c.\right)\\
&-\frac{1}{2}\sum_{m=-L/2}^{L/2}\sum_{m' = -L/2+1}^{L/2-1} \sum_{k=1}^{L}\sum_{\tilde{m}'=-L/2}^{L/2}\sum_{\tilde{k}=1}^{L}\frac{1}{-U_{\Uparrow\Downarrow}+ \Omega(m-m')} \left(\ket{\tilde{m}'\tilde{k}}\bra{m'k} \> \bra{\tilde{m}'\tilde{k}}\hat{H}_{J}\ket{m}\bra{m}\hat{H}_{J} \ket{m'k} +h.c.\right).
\end{aligned}
\end{equation}
\normalsize
To evaluate these, we need the actual matrix elements of individual spin operators $\hat{\sigma}_{k_X,k_Y}^{x}$ of which the perturbation consists. These are given by~\cite{swallows2011spinwaveMatrixElements}
\begin{equation}
\bra{m}\hat{\sigma}_{n}^{'x}\ket{m'k} = \frac{e^{2\pi i k n/ L}}{2} \left(-\sqrt{\frac{\left(\frac{L}{2}+m\right)\left(\frac{L}{2}+m-1\right)}{L^2(L-1)}}\delta_{m,m'+1} + \sqrt{\frac{\left(\frac{L}{2}-m\right)\left(\frac{L}{2}-m-1\right)}{L^2(L-1)}}\delta_{m,m'-1}\right),
\end{equation}
where $n \in 1 \dots L$ labels all of the spins using whatever indexing is convenient. For example, since we are working in 2D we write $n = L_X ( \frac{L_Y}{2\pi} k_Y - 1) + \frac{L_X}{2\pi}k_X$ for conventional quasimomenta $k_X \in \frac{2\pi}{L_X}\{1,2,\dots,L_X\}$ and $k_Y \in \frac{2\pi}{L_Y} \{1,2,\dots L_Y\}$. Inserting these matrix elements into the sums above leads to
\footnotesize
\begin{equation}
\begin{aligned}
\hat{H}_{J,\mathrm{eff}}=&-\frac{2(J_0 + J_1)^2 U_{\Uparrow\Downarrow}}{(L-1)(U_{\Uparrow\Downarrow}^2-\Omega^2)}\sum_{m=-L/2}^{L/2}m^2 \ket{m}\bra{m}\\
&+\frac{(J_0 + J_1)^2 U_{\Uparrow\Downarrow}}{(L-1)(U_{\Uparrow\Downarrow}^2 - \Omega^2)}\sum_{m=-L/2}^{L/2-2} \sqrt{\left(\frac{L}{2}+m+2\right)\left(\frac{L}{2}+m+1\right)\left(\frac{L}{2}-m\right)\left(\frac{L}{2}-m-1\right)}\left(\ket{m+2}\bra{m} + h.c.\right)\\
&-\frac{2(J_0+J_1)^2\Omega}{U_{\Uparrow\Downarrow}^2 -\Omega^2} \sum_{m=-L/2}^{L/2} m \ket{m}\bra{m}.
\end{aligned}
\end{equation}
\normalsize
These sums can all be identified with simple collective-spin terms. The first two lines correspond to a twisting term $\hat{S}^{x}\hat{S}^{x}$, while the last line is a single particle term. We thus get
\begin{equation}
\hat{H}_{J,\mathrm{eff}} = \frac{1}{L-1}\frac{4(J_0+J_1)^2 U_{\Uparrow\Downarrow}}{U_{\Uparrow\Downarrow}^2 - \Omega^2}\hat{S}^{x}\hat{S}^{x} - \frac{2(J_0+J_1)^2 \Omega}{U_{\Uparrow\Downarrow}^2 - \Omega^2}\hat{S}^{z}.
\end{equation}
This is the second-order perturbative contribution of the tunneling to the model's time-evolution. The first piece is a twisting term, while the second piece is just a correction to the diagonal single-particle drive frequency. Note that the single-particle correction is negligible compared to the bare magnitude of the drive $\Omega$, and we will neglect it, keeping only the twisting term.

\subsection{Full collective model}
We have now evaluated the zeroth-order contributions from the nearest-neighbour interactions $\hat{H}_{\mathrm{int}}^{\mathrm{NN}}$ and the perturbative, second-order contributions from tunneling $\hat{H}_{J,\mathrm{eff}}$. Adding these to the unperturbed $\hat{H}_{0}$ yields the following collective model:
\begin{equation}
\begin{aligned}
\hat{H}_{\mathrm{OAT}} &= -\frac{U_{\Uparrow\Downarrow}}{L} \vec{S} \cdot \vec{S} - 2\chi \hat{S}^{x}\hat{S}^{x} + \Omega\hat{S}^{z},\\
\chi &= \frac{1}{L-1}\frac{2(J_0+J_1)^2 U_{\Uparrow\Downarrow}}{\Omega^2 - U_{\Uparrow\Downarrow}^2} - \frac{1}{L}\left(V_{ee}- 2V_{\Uparrow\Downarrow}\right),
\end{aligned}
\end{equation}
with $\chi$ same as in the main text. This is a "twist-and-turn" model containing a one-axis twisting term, and a drive that it does not commute with. However, since the drive term $\sim \Omega \hat{S}^{z}$ is much stronger than the squeezing rate $\chi$, we can make a rotating-wave approximation,
\begin{equation}
-2\chi\hat{S}^{x}\hat{S}^{x} \approx -2\chi \left[\frac{1}{2} \left(\hat{S}^{+}\hat{S}^{-} + h.c.\right)\right] = -\chi\vec{S}\cdot\vec{S} +\chi \hat{S}^{z}\hat{S}^{z}.
\end{equation}
The $-\chi\vec{S}\cdot\vec{S}$ portion of this is small compared with the bare $\frac{U_{\Uparrow\Downarrow}}{L}\vec{S}\cdot \vec{S}$ term and can be neglected. Note that any $\vec{S}\cdot \vec{S}$ term amounts to a constant energy shift when within the Dicke manifold anyways, and we only keep it in the Hamiltonian to emphasize the gap protection preventing leakage into other manifolds. This leaves us with
\begin{equation}
\hat{H}_{\mathrm{OAT}} \approx - \frac{U_{\Uparrow\Downarrow}}{L}\vec{S} \cdot \vec{S} +  \chi \hat{S}^{z}\hat{S}^{z} + \Omega\hat{S}^{z},
\end{equation}
which is the one-axis twisting model $\hat{H}_{\mathrm{OAT}}$ in the main text.
\end{appendices}

\end{document}